%% file: 0-paper.tex
\documentclass{sigchi}
\usepackage{balance, graphicx}
\usepackage[T1]{fontenc}
\usepackage{txfonts, mathptmx}
\usepackage[pdftex]{hyperref}
\usepackage{color, booktabs, textcomp}
\usepackage{microtype, ccicons, tabularx}
\usepackage{todonotes, enumerate}
\usepackage{csquotes}
\usepackage{rotating, datetime}
\usepackage{placeins}
\usepackage{parnotes}
\usepackage{enumitem}
\usepackage{calc}

\makeatletter
\g@addto@macro{\UrlBreaks}{\UrlOrds}
\makeatother

\input{z-commands}
\input{abstractAndTitle}

\def\emptyauthor{}

\definecolor{linkColor}{RGB}{6,125,233}
\hypersetup{%
  pdftitle={\plaintitle},
  pdfauthor={\emptyauthor},
  pdfkeywords={\plainkeywords},
  bookmarksnumbered,
  pdfstartview={FitH},
  colorlinks,
  citecolor=black,
  filecolor=black,
  linkcolor=black,
  urlcolor=linkColor,
  breaklinks=true,%
}

\begin{document}

\title{\plaintitle}


\numberofauthors{1}
\author{%
  \alignauthor{Sean Penney, Jonathan Dodge, Claudia Hilderbrand, Andrew Anderson, Logan Simpson, Margaret Burnett}\\
    \affaddr{Oregon State University}\\
    \affaddr{Corvallis, OR; USA}\\
    \email{ $\lbrace$ penneys, dodgej,  minic, anderan2, burnett $\rbrace$@oregonstate.edu }\\
}

\maketitle

\begin{abstract}
\abstractText
\end{abstract}



\input{1-Introduction}
\input{2-Background}

\input{3-Methodology}
\countedSection{Results}
\input{4-ResultsLimDey}

\input{5-ResultsPaths}
\input{6-ResultsDecisionPoints}
\input{8-Discussion}

\input{9-Conclusion}
\bibliographystyle{SIGCHI-Reference-Format}

\pagestyle{plain}

\balance{}
\bibliography{references}
\end{document}

%% file: z-commands.tex
\newcommand{\boldify}[1]{}
\newcommand{\FIXME}[1]{}

\newcommand{\pair}[2]{{%
Pair#1-P#2}}

\newcommand{\Ccode}[1]{{%
\textsl{#1}}}

\newcommand{\LimDey}[1]{{%
\texttt{#1}}}
    
\newcommand{\ValueWhy}[1]{{%
\textsl{#1}}}

\newcommand{\quotateInset}[3]{%
\vspace{-5pt}\begin{displayquote}%
\pair{#1}{#2}: ``\emph{#3}''%
\end{displayquote}}

\newcommand{\quotate}[3]{{%
\pair{#1}{#2}: ``\emph{#3}''}}

\newcommand{\quotateUnattrib}[1]{{%
``\emph{#1}''}}

\newcommand{\quotateNE}[1]{%
\vspace{-5pt}\begin{displayquote}%
Shoutcaster for Nerchio v Elazer: ``\emph{#1}''%
\end{displayquote}}

\newcommand{\quotateBI}[1]{%
\vspace{-5pt}\begin{displayquote}%
Shoutcaster for Byun v Iasonu: ``\emph{#1}''%
\end{displayquote}}

\def\ImplicationsText{Implications for a Future Interactive Explanation System}

\def\sc{StarCraft}

\newcommand{\REDACTOSU}[1]{[a U.S. university]}

\newcommand{\countedSection}[1]{\section{#1}\stepcounter{section}}
\newcommand{\countedSubsection}[1]{\subsection{#1}\stepcounter{subsection}}

\makeatletter
\def\url@leostyle{%
  \@ifundefined{selectfont}{
    \def\UrlFont{\sf}
  }{
    \def\UrlFont{\small\bf\ttfamily}
  }}
\makeatother
\def\pprw{8.5in}
\def\pprh{11in}

%% file: abstractAndTitle.tex
\def\abstractText{
Assessing and understanding intelligent agents is a difficult task for users that lack an AI background.
A relatively new area, called ``Explainable AI,'' is emerging to help address this problem, but little is known about how users would forage through information an explanation system might offer. 
To inform the development of Explainable AI systems, we conducted a formative study --- using the lens of Information Foraging Theory --- into how experienced users foraged in the domain of StarCraft to assess an agent.
Our results showed that participants faced difficult foraging problems.
These foraging problems caused participants to entirely miss events that were important to them, reluctantly choose to ignore actions they did not want to ignore, and bear high cognitive, navigation, and information costs to access the information they needed.

}

\def\plaintitle{
Toward Foraging for Understanding of StarCraft Agents:\\ An Empirical Study
}

\def\plainkeywords{
Intelligent Agents; 
Explainable AI; 
Intelligibility;
Content Analysis; 
Video Games; 
StarCraft; 
Information Foraging
}

%% file: 1-Introduction.tex
\countedSection{Introduction}

\boldify{Real-time strategy games are useful for AI research}

Real-time strategy (RTS) games are a popular test bed for artificial intelligence (AI) research, and platforms supporting such research continue to improve (e.g., ~\cite{vinyals}).
The RTS domain is challenging for AI due to real-time adversarial planning requirements within sequential, dynamic, and partially observable environments~\cite{ontanon}.
Since these constraints transfer to the real world, improvements in RTS agents can be applied to other domains, for example, mission planning and execution for AI systems trained to control a fleet of unmanned aerial vehicles (UAVs) in simulated environments~\cite{sycara2015abstraction}.
However, the intersection of two complex domains, such as AI and flight, poses challenges:
who is qualified to assess behaviors of such a system?
For example, how can a domain expert, such as a flight specialist, assess whether the system is making its decisions \emph{for the right reasons}?

\boldify{But, the inner workings of the AI systems aren't easily understood by people such as domain experts}

If a domain expert making such assessments is not an expert in the complex AI models the system is using, there is a gap between the knowledge they need to make such assessments vs. the knowledge they have in the domain.
To close this gap, a growing area known as ``Explainable AI'' aims to enable domain experts to understand complex AI system by requesting explanations.
Prior work has shown that such explanations can improve mental models~\cite{kulesza2015principles,kulesza2010explanatory}, user satisfaction~\cite{kapoor2010interactive}, and users' ability to effectively control the system~\cite{beltran2017don, bostandjiev2012tasteweights,kulesza2012tell}.

\input{figures/FigScreenshot}

\boldify{But there's not much known about how to help such people. So let's look at how experienced players (one kind of domain expert) go about figuring out why some other player or agent is doing whatever they're doing? }

However, little is known about what an RTS domain expert's information needs are -- what they need to have explained, in what sequence, and at what cognitive and time costs.  
Therefore, to inform explanation systems in this area, we conducted a formative study of how experienced RTS players would go about trying to understand and assess an intelligent agent playing the RTS game of StarCraft.

\boldify{Our setting was...}

Our setting was StarCraft replay files.
A StarCraft replay file contains an action history of a game, but no information about the players (i.e., no pictures of players and no voice audio).
This anonymized set-up enabled us to tell our participants that one of the players was an AI agent.
(We detail this design further in the Methodology section.)
In addition, the participants had functionality to seek additional information about the replay, such as navigating around the game map, drilling down into production information, pausing, rewinding, fast-forwarding, and so on (Figure~\ref{fig:screenshot}).

\boldify{And let's do our investigation using abstractions that may tell us how to better support them beyond the particular game of Starcraft. To understand their behaviors, we'll use IFT}

However, we wanted a higher level of abstraction than features specific to StarCraft.
Specifically, we aimed for (1)~applicability to other RTS environments, and (2)~connection with other research about humans seeking information.
To that end, we turned to Information Foraging Theory (IFT). 

IFT has a long history of revealing useful and usable information functionalities in other information-rich domains, especially web environments (e.g., \cite{pirolli2007information}) and software development environments (e.g., \cite{fleming2013information, piorkowski2015fix}).
Originally based on classic predator-prey models in the wild, its basic constructs are the \emph{predator} (information seekers like our participants) seeking \emph{prey} (information goals) along pathways marked by \emph{cues} (signposts) in an \emph{information environment} (such as the StarCraft replay environment).
The predator decides which paths to navigate by weighing the expected cost of navigating the path against the expected value of the location to which it leads.

\boldify{Now you know the big picture, what are the RQs?}
Drawing from this theory, we framed our investigation using the following research questions (RQs):
\vspace{-5pt}
\begin{enumerate}[labelindent=20pt,labelwidth=\widthof{\ref{last-item}},label=\arabic*.,itemindent=1em,leftmargin=!]
\item[\textbf{RQ1}] \emph{The Prey}: What kind of information do domain experts seek, how do they ask about it, and for what reasons?
\vspace{-5pt}
\item[\textbf{RQ2}] \emph{The Foraging Paths}: What paths do domain experts follow in seeking their prey, why, and at what cost?
\vspace{-5pt}
\item[\textbf{RQ3}] \emph{The Decisions and the Cues}: What decision points do domain experts consider to be most critical, and what cues lead them astray from these decision points? \label{last-item}
\end{enumerate}
\vspace{-5pt}

%% file: figures/FigScreenshot.tex
\begin{figure}[b!]
	\centering
    \vspace{-10pt}
	\includegraphics[width=.9\linewidth]{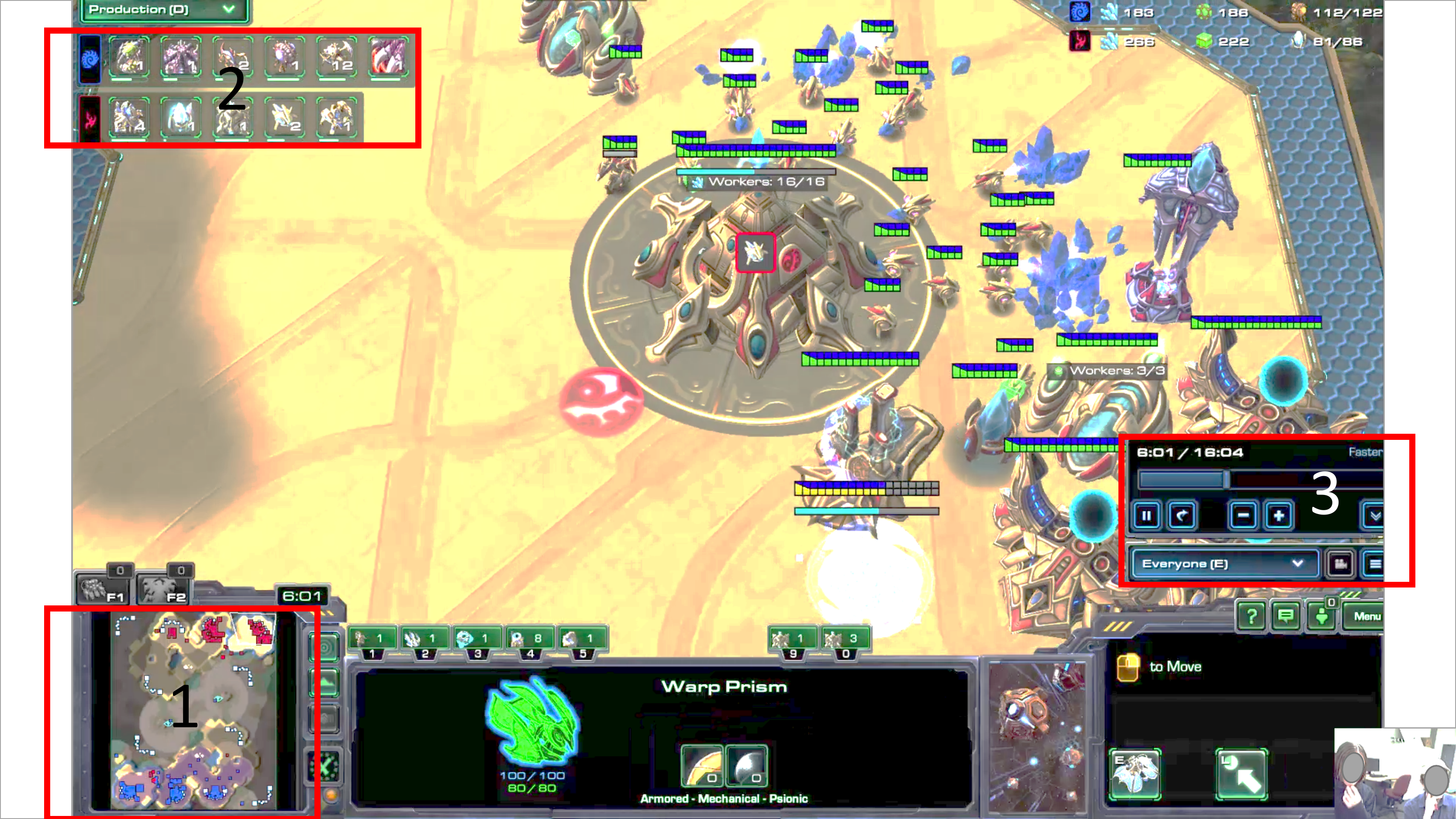}
	\caption{
    	A screenshot from our study, with participants anonymized (bottom right corner).
         Some important regions are marked with red boxes, including: (1: bottom left) The \emph{Minimap} offers a birds-eye view enabling participants navigate around the game map.
       (2: top left) Participants can use a drop-down menu to display the \emph{Production tab} for a summary of the build actions currently in progress.
       (3, middle right) \emph{Time Controls} allow participants to rewind/fast forward, change the speed. 
	}
	\label{fig:screenshot}
\end{figure}

%% file: 2-Background.tex
\countedSection{Background and Related Work}

\boldify{Why do we care about any of this? explanation helps mental models}

When assessing whether an AI agent is making its decisions for the right reasons, humans automatically develop mental models of the system~\cite{norman1983some}.
Mental models, defined as ``internal representations that people build based on their experiences in the real world,'' enable users to predict system behavior~\cite{norman1983some}.

\boldify{Explanations about system behavior improve mental models and satisfaction.}

Ideally, mental models of a system would help people gain the understanding they need to assess an AI agent, but this is not always the case.
Tullio et al.~\cite{tullio2007} examined mental models for a system that predicted the interruptibility of their managers.
They found that the overall structure of their participants' mental models was largely unchanged over the 6 week study, although they did discount some initial misconceptions.
However, their study did not deeply engage in explanation; it was mostly visualization.
In other work, Bostandjiev et al.~\cite{bostandjiev2012tasteweights} studied a music recommendation system and found that explanation led to a remarkable increase in user-satisfaction.
In an effort to improve mental models by increasing  transparency of a machine learning system, Kulesza et al.~\cite{kulesza2015principles} identified principles for explaining (in a ``white box'' fashion) how a machine learning based system makes its predictions more transparent to the user.
In their study, participants using a prototype following the principles observed an improvement of their mental model quality by up to 52\%.

\boldify{Further, explanation has been shown to improve users' ability to control the system.}

Several studies have also found that explanations have been able to improve users' ability to actually \emph{control} the system.
Stumpf et al.~\cite{stumpf2007} investigated how users responded to explanations of machine learning predictions, finding that participants were willing to provide a wide range of feedback in an effort to improve the system.
Kulesza et al.~\cite{kulesza2012tell} found that the participants who were best able to customize recommendations were the ones who had adjusted their mental models the most in response to explanations about the recommender system.
Further, those same participants found debugging more worthwhile and engaging.
Kapoor et al.~\cite{kapoor2010interactive} found that interacting with explanations enabled users to construct classifiers that were more aligned with target preferences, alongside increased satisfaction.
Beltran et al.~\cite{beltran2017don} presented a novel gestural approach to querying text databases,
allowing users to refine queries by providing reasons why the result was correct or incorrect. 
Their results indicated that action explanation allowed for more efficient query refinement.

\boldify{In the domain of RTS, has any of this been studied? Well, a little...}

However, in the domain of intelligent agents in RTS games, although there is research into AI approaches~\cite{ontanon}, there is only a little research investigating what \emph{humans} need or want explained.
Cheung et al.~\cite{Cheung:2011:SSU:1978942.1979053} studied how people watch the RTS genre, creating personas for various types of viewers.
Metoyer et al.~\cite{metoyer2010explaining} studied how experienced players explained the RTS domain to novice users while demonstrating how to play the game.
Finally, McGregor et al.'s~\cite{mcgregor2015mdp} work is also pertinent, describing support for testing and debugging in settings with thousands of decisions being made sequentially.
The work most similar to our own is Kim et al.'s~\cite{kim2016evaluation} study of intelligent agent assessment in StarCraft.
Their study invited experienced players to assess skill levels and overall performance of AI bots by playing against them.
They observed that the humans' ranking differed from an empirical ranking based on the bots' win rate at AI competitions.
Our study differs from theirs in that our participants did not play, but instead strove to \emph{understand and explain} by interacting with a game replay.

\boldify{But how do we access Explanations? via questions which Lim-Dey have categorized into various intelligibility types, and characterized in terms of DEMAND}

In everyday conversation, people obtain explanations by asking questions.
Drawing upon this point, Lim et al.~\cite{lim2009} have categorized questions people ask about AI systems in terms of ``intelligibility types.'' 
In their work, they investigated participants' information demands about context-aware intelligent systems powered by decision trees, determining which explanation types provided the most benefit to users.
They found the most often demanded questions were \LimDey{Why} and \LimDey{Why not} (Why did or didn't the system do X?).
We provide more details of that work and build upon it in when we discuss RQ1 results.

\boldify{Previous research has largely focused on SUPPLYING why/why not answers}

In recognition of the particular importance of these two types of questions, researchers have been working on \LimDey{Why} and \LimDey{Why not} explanations in domains such as database queries~\cite{bhowmick2013not, he2014answering}, robotics~\cite{hayes2017improving, lomas2012robots, Rosenthal2016robots}, email classification~\cite{kulesza2011oriented}, and pervasive computing~\cite{vermeulen2010pervasivecrystal}.
These types of explanations have also attracted attention from the social sciences, which seek to help ground AI researchers' efforts in cognitive theories~\cite{miller2017social}.
Other research has demonstrated that the intelligibility type(s) the system supported impact which aspects of users' attitudes are affected. 
For example, Cotter et al.~\cite{cotter2017explaining} found that justifying
\emph{why} an algorithm works the way it does (but not  \emph{how} it works) increased users' confidence (blind faith) in the system --- but not for improving their \emph{trust} (beliefs which inform a full cost-benefit analysis) in the system.
Further, it seems that the relative importance of the intelligibility types may vary from one domain to another.
For example, Castelli et al.~\cite{castelli2017happened} found that in the smart homes domain, users showed a strong interest in \LimDey{What} questions, but few of the other intelligibility types.

\boldify{How can we study information sought in our domain?: Enter IFT, But we should make it clear that IFT is based on individual perception of unknown quantities}

We drew upon Information Foraging Theory (IFT) to investigate the information that people would seek in the RTS domain.
In IFT terms, when deciding where to forage for information, predators (our participants) make cost/benefit estimates, weighing the information value per time cost of staying in the current \emph{patch} (location on the game map or tab with supplemental information) versus navigating to another patch~\cite{pirolli2007information}.  
However, predators are not omniscient: they decide based on their \emph{perceptions} of the cost and value of the available options. 
Predators form these perceptions using their prior experience with similar patches~\cite{piorkowski2015fix} and the \emph{cues} (signposts in their information environment like links and indicators) that point toward various patches.
Of course, predators' perceived values and costs are often inaccurate~\cite{piorkowski2016foraging}.

\boldify{IFT has been used in many domains but, to our knowledge, not in RTS environments.}

IFT constructs have been used to understand humans' information-seeking behavior in other domains, particularly web navigation~\cite{chi2001using, fu2007snif},
debugging~\cite{fleming2013information, kuttal2013predator, piorkowski2015fix},
and other software development tasks~\cite{niu2013departures, perez2014diagnosis, piorkowski2016foraging, srinivasa2016foraging}.
However, to our knowledge, it has not been used in RTS environments like StarCraft.
Our paper aims to help fill this gap.

%% file: 3-Methodology.tex
\countedSection{Methodology}

\boldify{Overall design: think aloud pair agent assessment}
We conducted a pair think-aloud study, where participants worked to understand and explain the behavior of an intelligent agent playing StarCraft II, a real-time strategy (RTS) game.
We used the pair think-aloud design to capture their ongoing efforts to understand the behaviors they witnessed. 

\boldify{Platform: Starcraft 2 match between pros}

StarCraft II is a popular RTS game ~\cite{ontanon} that has been used for AI research~\cite{vinyals}. 
The particular match\footnote{We used game 3 of the match between professional players \emph{ByuL} and \emph{Stats} during the \emph{IEM Season XI - Gyeonggi} tournament.
The IEM tournament series is denoted as a ``Premier Tournament,'' by TeamLiquid, a multi-regional eSports organization that takes a keen interest in professional StarCraft II.
The replay file is public at: \url{http://lotv.spawningtool.com/23979/}}
we used featured professional players and was part of a top level tournament.
The replay we chose to analyze was a representative sample in terms of game flow, e.g., initially building up economy, some scouting, transitioning to increasing combat~\cite{ontanon}.

\input{tables/TableParticipants}

\boldify{We used deception so they would think they were assessing an AI}

Because we were interested in how participants would go about understanding an intelligent agent's behaviors, we hid the players' names, instead displaying them as \emph{Human} and \emph{CPU1}, and told participants that one of the players was under AI control --- even though that was untrue. 
To encourage them to aim for a real understanding of an agent that might have weaknesses, we also told them the AI was not fully developed and had some flaws. 
Participants were generally convinced that the player was an AI.
For example, \pair{5}{10} speculated about the implementation, 
\quotateUnattrib{he must have been programmed to spam.}
Participants did notice, however, the AI at times behaved like a human:
\quotateInset{10}{20}{Okay, I've not thought of that angle for some reason: The AI trying to act like a human.}

\boldify{deception was necessary because...}

Instead of using deception to simulate an intelligent agent with a human player, an alternative design might use a replay of a game with an intelligent agent playing.
However, we needed replay files with both interactive replay instrumentation and high-quality gameplay.
We were unable to locate an intelligent agent in the RTS domain with high enough quality for our investigation, i.e., without limitations like exploiting ``a strategy only successful against AI bots but not humans''~\cite{kim2016evaluation}.

\countedSubsection{Participants}

\boldify{Potential participants had to satisfy the following criteria...}

We wanted participants familiar with the StarCraft user interface and basic game elements, but without knowledge of machine learning or AI concepts, so we recruited StarCraft players at \REDACTOSU{Oregon State University}  with at least 10 hours of prior experience -- but excluding computer science students.
Also, to avoid language difficulties interfering with the think-aloud data, we accepted only participants with English as their primary language.
As per these criteria, 20 undergraduate students participated (3 females and 17 males), with ages ranging from 19--41, whom we paired based on availability.
Participants had an average of 93 hours of casual StarCraft experience and 47 hours of competitive StarCraft experience (Table~\ref{table:participants}).

\countedSubsection{Procedures}

\subsubsection{Main Task's Procedures}

\boldify{Groups of 2 participants replayed game and comment/observe while having the audio and video be recorded}

For the main task, each pair of participants interacted with a 16-minute StarCraft II replay while we video-recorded them.
The interactive replay instrumentation, shown in Figure~\ref{fig:screenshot}, allowed participants to actively forage for information within the replay, and we gave them a short tutorial of its capabilities. 
Examples of ways they could forage in this environment were to move around the game map, move forward or backward in time, find out how many units each player possessed, and drill down into specific buildings or units. 

\boldify{They pair-watched and wrote things down}

Participants watched and foraged together as a pair to try to make sense of the agent's decisions.
One participant controlled the keyboard and mouse for the first half of the replay, and they switched for the second half.
To help them focus on the decisions, we asked them to write down \emph{key decision points}, which we defined for them as, ``an event which is critically important to the outcome of the game.'' 
Whenever they encountered what they thought was a key decision point, they were instructed to fill out a form with its time stamp, a note about it, and which player(s) the decision point was about. 







\input{tables/TableInterviewQuestions}

\subsubsection{Retrospective Interview's Procedures}

\boldify{Last, we conducted a post task interview asking participants various questions about specific actions they made during the main task.}

After the main task, we conducted a multi-stage interview based on the actions the participants took during the main task.
To add context to what participants wrote down during the main task, we played parts of our recording of their main task session, pausing along the way to ask why they chose the decision points they did. 
The wording we used was: ``\emph{In what way(s) is this an important decision point in the game?}''



\boldify{interview stage 3: navigations, sampling}

We went through the main task recording again, pausing at their navigations to ask the questions in Table~\ref{table:interviewQuestions}.
Since there were too many to ask about them all, we sampled pre-determined time intervals to enable covering several instances of each type of navigation for all participant pairs.




\countedSubsection{Analysis Methods}

\boldify{We used Lim Dey coding to answer RQ1.  Our researchers had previously achieved IRR on another corpus, which we argue carries over to this corpus.}

To answer RQ1, we qualitatively coded instances in the main task where participants asked a question out loud, using the code set outlined later in Table \ref{table:limDeyCoding}.
Our researchers had used this code set on a different corpus [Removed for anonymized review] in which they independently coded 34\% of the corpus and achieved 80\% inter-rater reliability (IRR).  
In this study, the same researchers who achieved this IRR split up the coding of the current data.

\boldify{Affinity diagramming and qualitative coding was performed to answer the question ``Why was the participant seeking information.''}

To answer RQ2, we qualitatively analyzed the participants' responses to the retrospective interview questions. 
To develop a code set to answer the question ``\emph{Why was the participant seeking information?}'' we started with affinity diagramming to generate groups of answers, which was performed by a group of four researchers. 
The affinity diagram led to the following codes: \ValueWhy{Monitoring State}, \ValueWhy{Updating Game State}, \ValueWhy{Obsolete Domain}, and \ValueWhy{New Event}, as shown later in Table~\ref{table:whyCoding}.
Two researchers then individually qualitatively coded the participants' responses using this code set on 20\% of the data.  
Given that our IRR on this portion was 80\%, one researcher then completed the rest of the coding alone.

\boldify{For RQ3, We examined key decision points users identified. Groups of decisions points were identified during an affinity diagramming session, and qualitative coding was then performed using the groups to guide the coding.}

To answer RQ3, we qualitatively coded the decision point forms the participants used during the main task.
Here again, we developed a code set using affinity diagramming.
The four higher level codes we used were building/producing, scouting, moving, and fighting.
We coded 24\% of the 228 identified decision points according to this code set, and reached IRR of 80\%, at which point one researcher coded the rest of the data.

%% file: tables/TableParticipants.tex
\begin{table}[t]
	\centering
    \small
    \begin{tabular}{@{}l|ll|ll|ll@{}}

\multicolumn{3}{l|}{} &
\multicolumn{2}{c|}{\emph{SC2 Hours}} &
\multicolumn{2}{c}{\emph{RTS Hours}} \\\cline{4-7}

	\multicolumn{3}{l|}{\textit{Participant: Age, Gender, Major}}

	&
	\textit{Casual}
    
    &
	\textit{Comp.}
  
    &
	\textit{Casual}
    
    &
	\textit{Comp.} 

\parnoteclear
      
        \\ \hline
		 \pair{1}{1}
         & 41 M 
         & EE\parnote{Electrical Engr.}
         & 200
         &100
         & 500
         &300
      
         \\
         \pair{1}{2}
         & 20 M  
         &  ECE\parnote{Electrical \& Computer Engr.}
         & 50
         &20
         & 30
         & 30
       
         \\ \hline
         \pair{2}{3}
         & 23 M  
         &  CE \parnote{Chemical Engr.}
         & 10& 5
         & 25& 55
       
                  \\
         \pair{2}{4}
         & 23 M  
         &  ME\parnote{Mechanical Engr.}
         & 100
         & 200
         & 50
         & 0
    
                           \\ \hline
         \pair{3}{5}
         & 21 M  
         & EE
         & 50& 0
         & 150& 12

                                    \\
         \pair{3}{6}
         & 27 M   
         & CE
         & 15& 2
         & 150& 10
         
                                    \\ \hline
         \pair{4}{7}
         & 23 M   
         & CE
         & 40& 20
         & 20& 30
        
                                    \\ 
         \pair{4}{8}
         & 28 F   
         & EnvE\parnote{Environmental Engr.}
         & 200& 100
         & 300& 30
      
                                    \\ \hline
         \pair{5}{9}
         & 21 M   
         & BE\parnote{Biological Engr.}
         & 40& 40
         &100& 0
        
                                    \\
         \pair{5}{10}
         & 19 M   
         & ECE
         & 700& 300
         & 50& 0
   
                                    \\ \hline
         \pair{6}{11}
         & 22 M   
         & BE
         & 100& 2
         &160& 100
         
                                    \\
         \pair{6}{12}
         & 22 F   
         & EnvE
         & 0& 70
         & 0& 0
         
                                    \\ \hline
         \pair{7}{13}
         & 22 M   
         & CE
         & 15& 60
         & 100& 50
         
                                    \\ 
         \pair{7}{14}
         & 20 M   
         & BE
         & 35& 3
         & 40& 0
 
                                    \\ \hline
         \pair{8}{15}
         & 23 M   
         & CE
         & 10& 0
         & 100& 5
      
                                    \\
         \pair{8}{16}
         & 22 M   
         & BE\parnote{Business Entrepreneurship}
         & 16& 1
         & 15& 0

                                             \\ \hline
         \pair{9}{17}
         & 21 M   
         & PS\parnote{Political Science}
         & 90& 5
         & 500& 80
        
                                             \\ 
         \pair{9}{18}
         & 19 M   
         & ME
         & 100& 0
	& 20& 0
    
                                             \\ \hline
         \pair{10}{19}
         & 24 F   
         & FA\parnote{Fine Arts}
         & 5& 5
         & 0& 0
   
                                             \\ 
         \pair{10}{20}
         & 23 M   
         & EdEn\parnote{Education \& English}
         & 80& 15
         & 50& 0

        \end{tabular}
    \normalsize
	\caption{Participant demographics and their casual vs. competitive (Comp.) experience. 
    SC2 is StarCraft II, and RTS is any other Real-Time Strategy game.
    }
    \vspace{-4pt}
    \parnotes
	\label{table:participants} 
    \vspace{-10pt}  	
\end{table}

%% file: tables/TableInterviewQuestions.tex
\begin{table}[t]
	\centering
    \small
	\begin{tabular}{@{}p{.95\linewidth}@{}}

\textbf{When a participant paused the replay...}\\
		\hspace{.025\linewidth} - What about that point in time made you stop there?
        \\
        \hspace{.025\linewidth} - Did you consider stopping on that object at any other point in time?
        \\\hline
        
\textbf{When a participant navigated to a patch...}\\
 		\hspace{.025\linewidth} - What about that part of the game interface/map made you click there?
        \\
        \hspace{.025\linewidth} - Did you consider clicking anywhere else on the game interface/map?
        \\\hline
 
\textbf{When a participant navigated away from a patch (or unpaused)...}\\
 		\hspace{.025\linewidth} - Did you find what you expected to find?
        \\
        \hspace{.025\linewidth} - What did you learn from that click/pause?
        \\
        \hspace{.025\linewidth} - Did you have a different goal for what to learn next?
	\end{tabular}
    \normalsize
	\caption{Interview questions (drawn from prior IFT research~\protect\cite{piorkowski2016foraging}), and the triggers that caused us to ask them.}
	\label{table:interviewQuestions}   	
\end{table}

%% file: 4-ResultsLimDey.tex
\countedSubsection{RQ1: The Prey}

\input{tables/TableLimDeyCoding}

\boldify{IFT tells us there are prey. We can learn about them thru their questions. }

To understand how predators seek prey in the RTS domain, we analyzed questions participants asked during the main task. 
To situate our investigation in the literature of humans trying to understand AI, we coded the utterances using the Lim \& Dey intelligibility types~\cite{lim2009assessing} (Table \ref{table:limDeyCoding}).

\boldify{We coded using Lim/Day. Participants predominantly asked \LimDey{What}~Lim \& Dey questions, and very few of the others. This contrasts with research from Lim \& Dey, in which \LimDey{Why}~was the most demanded explanation type from users.}

The results were surprising.  
Although prior research has reported \LimDey{Why} questions to be much in demand \cite{lim2009assessing,lim2009},
only 10\% of our participants' questions fell into the \LimDey{Why did}~and \LimDey{Why Didn't}~categories combined (Table \ref{table:limDeyBySession}).
Over 70\% of our participants' questions pertained to \LimDey{What}.

\boldify{let's look at what experts supply (shoutcasters' expls) vs demand (participants' qus) to get some insights into this. (Co-authors: of course, we know it matches from our CHI submission, but we can't cite that yet...) }

To get a sense of how representative our participants' questions were, we turned to the experts --- namely, professional explainers in this domain, known as ``shoutcasters.''
Because we were interested in the very best explainers in this domain, we restricted our search for shoutcaster videos to those that fit the description in the Methodology (top level tournament, professional players).
From this pool, we used the same code set as Table \ref{table:limDeyCoding} to analyze two professionally explained games: Byun vs. Iasonu\footnote{Game 2 of \emph{Byun} vs. \emph{Iasonu} in the \emph{2016 IEM Gyeonggi} tournament, available at: \url{https://sc2casts.com/cast20681-Byun-vs-Iasonu-BO3-in-1-video-2016-IEM-Gyeonggi-Group-Stage}}
and Nerchio vs. Elazer\footnote{Game 2 of \emph{Nerchio} vs. \emph{Elazer} in the \emph{2016 WCS Global Finals} tournament, available at: \url{https://sc2casts.com/cast20439-Nerchio-vs-Elazer-BO3-in-1-video-2016-WCS-Global-Finals-Group-Stage}}.

Consistent with our participants' questions, the shoutcasters' explanations were dominated by answers to \LimDey{What} questions.
In the Nerchio vs. Elazer game, shoutcasters answered  \LimDey{What}~questions 54\% of the time, and in Byun vs. Iasonu they provided \LimDey{What}~answers 48\% of the time.
Since shoutcasters are hired to provide what game audiences want to know, their consistency with our participants' questions suggest that this distribution of questions was typical for the domain.

\input{tables/TableLimDeyBySession}

\subsubsection{The many flavors of ``What'' prey}

\boldify{Hmmm, maybe all they need is a play-by-play?}

Why such a difference from prior research results?
One hypothesis is that, in this kind of situation, participants' prey was simply ``play-by-play'' information. 
However, this hypothesis is not well supported by the data.
Although participants did seek \emph{some} play-by-play information (\quotate{3}{5}{...so he just killed a scout, right?}), several common prey patterns in their \LimDey{What}~questions went beyond play-by-play.
Three of these patterns accounted for about one-third of the \LimDey{What}~questions.

\boldify{Why were they asking "What"? (Case 1) Drilling down, in order to figure out what units an agent was actively building or has etc.  Case 1:  drilling down, which sometimes was costly. Also, we noticed that Shoutcaster explanations answered these types of questions.}

\emph{The ``drill-down What'' of current state}: One common question type participants asked when pursuing prey were questions that involved drilling down to find the desired information. 
Half of the pairs 
asked drill-down \LimDey{What}~questions about the game players' unit production or composition. 
There were 21 instances of this type of \LimDey{What} question alone, accounting for almost 15\% of the total \LimDey{What} questions.
For example, the following question required the participants to drill down into several structures on the map to answer it:
\quotateInset{3}{6}{Is the human building any new stuff now?}

Navigating in pursuit of this kind of prey was often costly. 
The least expensive way was navigating via a drop-down menu (2 clicks) in region 2 of Figure \ref{fig:screenshot}, but participants instead often foraged in other ways.
For example, to find the answer to a question (like \pair{3}{6}'s earlier) participants sometimes navigated to several unit producing structures on the map, into a structure, and then on to the next.
For example, Pair 3 made seven navigations to answer their question about ``building new stuff.''

Shoutcasters' comments closely matched the participants' interest in drilling down: 18\% of shoutcasters' \LimDey{What} comments answered drill-down questions, compared to the 15\% of our participants' drill-down \LimDey{What}s.
As an example of the match to shoutcasters' comments, \pair{6}{12}'s drill-down question about unit composition: \quotateUnattrib{I think, well, we have a varied composition, besides roaches, and what are these?}
would be well-matched to shoutcaster explanations such as these:
\quotateBI{[the player has] 41 zerglings at the moment.} 
\quotateNE{And 12 lings as well, [and] that's just a few more lings than you normally see.}
This suggests using shoutcasters as a possible content model for future explanation systems, given that shoutcasters' ``supply'' of explanations seem to match well with participants' ``demand'' for explanations of this type.

\boldify{Case (2) Temporal: Users also asked \LimDey{What} questions to fill in gaps in temporal knowledge about game events.}

\emph{The ``temporal What'' of past states}: A second common prey pattern, used by almost half (4 of the 10) participant pairs, was asking \LimDey{What}~questions to fill in gaps regarding past states.
These accounted for 15 instances (about 10\%) of their \LimDey{What}s.
\quotateInset{3}{6}{When did he start building [a] robotics facility?}

As in the drill-down pattern, the participants' demand for temporal \LimDey{What}s matched well with the shoutcasters' supply: about 10\% of the shoutcasters' \LimDey{What}  explanations reminded listeners of some past event pertinent to current game state.
For example:
\quotateBI{...the plus 1 carapace early upgrade ...[is] actually paying off.}


\boldify{Case (3): A step up: There were also \LimDey{What} questions due to questions about what is going on. These seem more difficult to answer.}

\emph{The ``higher-level What''}: 
The third common prey pattern was at a higher level of abstraction than the specific units or events, aiming instead toward more general understanding of what was going on in the game.
These \LimDey{What}~questions arose 12 times (about 8\%) across all instances.
For example, \pair{10}{20} asked, \quotateUnattrib{What's going on over there?} in which ``there'' referred to a location on the map with military units that could have been gearing up for combat. 
The shoutcasters seemed enthusiastic about providing this kind of information\footnote{We did not count the number of shoutcaster comments that answered this question because we could not narrow them down in this way. That is, although many of their comments \emph{could} be said to be applicable to this type of question, the same comments were also applicable to more specific questions.}, perhaps because it provided opportunities to add nuance and insight to their commentary.
For example: 
\quotateBI{This is about to get crazy because [of] this drop coming into the main base [and] the banelings trying to get some connections in the middle.}
\quotateNE{I like Elazer's position; he's bringing in other units in from the back as well.}

\subsubsection{Questioning the unexpected}

\boldify{This brings us to violated expectations -- What meets Why...}

Lim \& Dey's reported that when a system behaved in unexpected ways, users' demand to know \LimDey{Why} increased~\cite{lim2009assessing}.
Consistent with this, when our participants saw what they expected to see, they did not ask \LimDey{Why} or \LimDey{Why-didn't} questions. 
For example, Pair 4 and Pair 5 did not ask any \LimDey{Why} or \LimDey{Why didn't} questions at all.
Instead, they made remarks like the following:  
\quotateInset {4}{7}{the Zerg is doing what they normally do.}
\quotateInset {4}{8}{[The agent is] kind of doing the standard things.}
\quotateInset {5}{10}{This is a standard build.}

However, in cases of the unexpected,
a fourth \LimDey{What} prey pattern arose, in which participants questioned the phenomena before them.
We counted 9 \LimDey{What}~questions of this type:
\quotateInset{9}{17}{...interesting that it's not even using those.}
\quotateInset{10}{19}{I don't get it, is he expanding?}
\quotateInset{10}{19}{Wow, what is happening? This is a weird little dance we're doing.}
\quotateInset{10}{20}{<when tracking military units> What the hell was that?}


\boldify{Why questions: According to Lim \& Dey, users demand less \LimDey {Why} explanations when expectations are not violated, and that seems to apply here.}

The unexpected also produced \LimDey{Why} questions.
About half of the participants' \LimDey{Why} and \LimDey{Why-Didn't} questions came from seeing something they had not expected or \emph{not} seeing something they had expected.
For example:
\quotateInset{1}{1}{<noticing a large group of units sitting in a corner> Why didn't they send the big army they had?}
\quotateInset {10}{19}{Oh, look at all these Overlords. Why do you need so many?}

\subsubsection{\ImplicationsText}

\boldify{ 
1. If shoutcasters are providing what users need ==> explanation systems can be designed based on shoutcasters (gold std). (But, we need to be careful! Within the categories of Lim and Dey questions, questions and answers can vary widely in many aspects. Furthermore, some of these may be  difficult for an explanation system to answer (What is going on type questions)). 1a. Over 70\% were WHAT questions.  Triangulates reasonably to shoutcasters' answers. ==> Expl systems should support those.}

Using the Lim \& Dey intelligibility types (What, Why, etc.) to categorize the kinds of prey our participants sought produced implications for shoutcasters as possible ``gold standards'' for informing the design of a future automated explanation system in this domain.  
For example, the high rate of \LimDey{What} questions from participants matched reasonably well with a high rate of \LimDey{What} answers from shoutcasters.
Drawing explanation system design ideas  from these  expert explainers may help inform the needed triggers and content of the system's \LimDey{What}~explanations.

\boldify{2. Three common WHAT prey patterns, covering about a third of the WHATs: (a) Drill-down (state I haven't seen but I need) (b) Temporal (state I have seen but I forget) (c) Higher-level (what's the point): ==> Expl systems should support those.}

Also, the dominance of \LimDey{What}~questions point to participants' prioritizing of state information in this domain.
Drill-down \LimDey{What}s were about state information they hadn't yet seen, temporal \LimDey{What}s were about past states they either hadn't seen or had forgotten, and higher-level \LimDey{What}s were about understanding the purpose of a current or emerging state.
Further, the shoutcasters matched and sometimes exceeded the participants' rate of \LimDey{What}s in each of these categories with their explanations.
This suggests that in the RTS domain, an explanation system's most sought-after explanations may be its explanations relating to state.   

\boldify{3. Some WHATs and lots of WHYs about violated expectations: seeing something unexpected or not seeing something expected.  ==> If an expl system can figure out what's "expected", it can better know what people will want explained (the things not in that set).}

Also, as noted in prior research, unexpected behaviors (or omissions of expected behaviors) led to increases in questions of both the \LimDey{What} and the \LimDey{Why} intelligibility types~\cite{lim2009assessing}.  
If an explanation system can recognize unexpected behavior, it could then better predict when users will want \LimDey{Why} and \LimDey{What} explanations to understand the deviation from typical behavior.

\boldify{4. Note that some very expensive (navigationally)! And we're about to find out even more about how expensive (cognitively) their foraging was next.}

Finally, the cost of navigating to some of the prey at times became expensive, which points to the need for explanation systems to keep an eye on the cost to users of obtaining that information.
In this section, this came out in the form of navigation actions.
The next section will point to costs to human cognition as well.



%% file: tables/TableLimDeyCoding.tex
\begin{table}
	\centering
    \small
	\begin{tabular}{@{}p{.89\linewidth}|l@{}}
		\emph{Intelligibility Type}&  	
        \textit{Freq}
        \\ \hline
		\LimDey{What}: 
        What the player did or anything about game state.
        \\
        \hspace{.025\linewidth} -\quotate{3}{5}{So he just killed a scout right?}
        & 148
		\\ \hline
        
		\LimDey{What-could-happen}: 
        What the player could have done or what will happen.
        \\
        \hspace{.025\linewidth} -\quotate{5}{10}{What's he gonna do in response?}
        & 16
        \\ \hline
        
        \LimDey{Why-did}: 
        Why the player performed an action.
        \\
        \hspace{.025\linewidth} -\quotate{10}{20}{What was the point of that?} 
        & 14
		\\ \hline
        
		\LimDey{How-to}: 
        Explaining rules, directives, audience tips, high level strategies.
        \\
        \hspace{.025\linewidth} -\quotate{3}{5}{You have to build a cybernetics core, right?} 
        & 9
		\\ \hline
        
		\LimDey{*How-good/bad-was-that-action}: 
  		 Evaluation of player actions.
        \\
        \hspace{.025\linewidth} -\quotate{10}{19}{Like, clearly it didn't work the first time, is it worth it to waste four units the second time?}
        & 8
        
		\\ \hline
        
		\LimDey{Why-didn't}:
        Why the player did not perform an action.
        \\
        \hspace{.025\linewidth} -\quotate{10}{20}{Why aren't they attacking the base?} 
        & 7
	\end{tabular}
    \normalsize
	\caption{Intelligibility type code set, frequency data, and examples.
    The code set is slightly modified (denoted by the asterisk) from the schema proposed by Lim \& Dey:
    We added \LimDey{How-good/bad-was-that-action} because the users wanted an evaluation of agent actions. }
	\label{table:limDeyCoding}   	
\end{table}

%% file: tables/TableLimDeyBySession.tex
\begin{table}
	\setlength{\tabcolsep}{3pt}
	\centering
    \small
	\begin{tabular}{@{}l|l|lllll|lllll@{}}
	     \textit{Question}
         & \textit{Total}
         & \rotatebox{90}{\textit{Pair 1}}
         & \rotatebox{90}{\textit{Pair 2}}
         & \rotatebox{90}{\textit{Pair 3}}
         & \rotatebox{90}{\textit{Pair 4}}
         & \rotatebox{90}{\textit{Pair 5}}
         & \rotatebox{90}{\textit{Pair 6}}
         & \rotatebox{90}{\textit{Pair 7}}
         & \rotatebox{90}{\textit{Pair 8}}
         & \rotatebox{90}{\textit{Pair 9}}
         & \rotatebox{90}{\textit{Pair 10}}\\ \hline
         
       \LimDey{What}
       & 148
       & 2
       & 3
       & 41
       & 1
       & 6
       & 14
       & 10
       & 1
       & 8
       & 62\\ \hline

       \begin{minipage}[t]{.325\columnwidth}
	\LimDey{What-could-happen}
	\end{minipage}
       & 16
       & 
       & 1
       & 1
       & 
       & 3
       & 1
       & 1
       & 
       & 2
       & 7\\\hline
       
       \LimDey{Why-did}
       & 14
       & 
       & 
       & 2
       & 
       & 
       & 3
       & 1
       & 
       & 
       & 8\\\hline
       
       \LimDey{How-to}
       & 9
       & 1
       & 
       & 3
       & 
       & 
       & 
       & 
       & 
       & 
       & 5\\ \hline
      
             \begin{minipage}[t]{.325\columnwidth}
	\LimDey{How-good/bad-was-}\\\LimDey{that-action}\vspace{2pt}
	\end{minipage}
       & 8
       & 
       & 
       & 3
       & 
       & 3
       & 
       & 
       & 
       & 
       & 2\\\hline
       
		\LimDey{Why-didn't}
       & 7
       & 1
       & 
       & 1
       & 
       & 
       & 
       & 
       & 
       & 
       & 5\\ \hline \hline
       
       {\hfill\textit{Total}}
       & 202
       & 4
       & 4
       & 51
       & 1
       & 12
       & 18
       & 12
       & 1
       & 10
       & 89\\

        \end{tabular}
        \normalsize
	\caption{Frequency of Lim \& Dey questions participants asked each other, by session.
    Note how often \LimDey{What} questions were asked, both by the population of participants as a whole, and by few individual pairs, where it was particularly prevalent.
    }    
	\label{table:limDeyBySession}   	
\end{table}

%% file: 5-ResultsPaths.tex
\countedSubsection{RQ2: The Foraging Paths}

\input{tables/TableWhyCoding}

\boldify{They had to deal with a lot of choices of which path to follow, and this got expensive for them cognitively. }

Various cognitive costs were incurred by participants by following paths to find prey.
As an information environment, RTS games have foraging characteristics that set them apart from other information environments that have previously been studied from an IFT perspective, such as web sites~\cite{pirolli2007information} and programming IDEs~\cite{piorkowski2016foraging}.
These previously studied domains are relatively static, with most changes occurring over longer periods.
In contrast, an RTS information environment changes rapidly and continually, driven by actions that do not originate from the foragers themselves.
As we will see, this caused participants to spend some time monitoring the overall game state, waiting for a suitable cue to appear for them to investigate further.

The \emph{number} of paths a forager might follow in an RTS information environment increases with the complexity of the game state, but path \emph{lengths} tend to be short.
This is conceptualized in Figure~\ref{fig:paths}.
This means that most questions are answered within a few navigations.
However, in foraging environments like IDEs, there might only be a few interesting links from any one information patch, but some can lead to  lengthy sequences of navigations (e.g., the ``Endless Paths'' problem~\cite{piorkowski2016foraging}).

\input{figures/FigPaths}

\subsubsection{Foraging in the RTS domain}

\boldify{How did this affect participants' foraging? let's look at the first minute to see what they look at if they have plenty of time to view everything of interest, versus what happened later.}

Interestingly, there was hardly any difference between RTS foraging and other environments at first.
During the early stages of a game, there are very few units, buildings, or explored regions for users to navigate to, so foraging is relatively straightforward.
As one participant put it:
\quotateInset{7}{14}{There is only so many places to click on at this point.}
As long as this remained the case, each relevant path could potentially be carefully pursued, similarly to an IDE.
Four participant pairs (2, 4, 7, 9) paused the replay for an average of 90 seconds within the first 1:30.
They studied individual objects and actions with a great deal of scrutiny, which was surprising considering the sparse environment.
In contrast, later on in the game, when 50 of the same unit existed, they received much less attention than when there was just one.

\input{tables/TableTaskTime}

\boldify{As game grows in complexity though, participants have to somehow evaluate MANY paths.}

Choosing among many available paths created cognitive challenges for participants. 
Participants needed to keep track of an increasing amount of information as the match progressed.
Each time a player performed an action, which added information, the participants could forage for this new information.
If the participant did so, we coded their navigation as a \ValueWhy{New Event}, which accounted for 26\% of our interviewed navigations (Table~\ref{table:whyCoding}).
For example: 
\quotateInset{10}{19}{...noticed movement in the Minimap, and that the Zerg troops were mobilizing in some fashion.  So I guess I just preemptively clicked...}

\boldify{Further, each action that creates paths may invalidate others, and they are updated OFTEN (due to high actions per minute (APM).}

The rate of path creation exacerbated the ``many paths'' problem.
Professional StarCraft players regularly exceed several hundred actions per minute (APM)~\cite{wong_2016}.
This meant that players performed rapid actions that changed the game state.
Each of these actions not only potentially created new paths;
they potentially updated the existing ones.
This caused the knowledge the participants had about paths that had not been recently checked to  become stale, which in turn led to a strong prevalence of two behaviors.
\ValueWhy{Update Game State} was very common in our data set, indicating that participants often needed to check on paths that may have been updated (21\% of interviewed navigations, Table~\ref{table:whyCoding}).
\quotateInset{8}{15}{...there's a big force again. Just checking it out to see if anything has progressed from earlier.}
\quotateInset{1}{2}{I was mainly just looking at the army composition, seeing how it had changed from the last fight, see if they had made any serious changes ...}

Note that this is slightly different from our second behavior, \ValueWhy{Monitoring State}, which is like updating game state, but with a nonspecific goal.
\ValueWhy{Monitoring State} was the most common reason for interviewed navigations (46\% of navigations were for the purpose of monitoring, Table~\ref{table:whyCoding}), for example:
\quotateInset{5}{9}{I noticed like the large mass of units on the map and I wanted to know what the player was doing with them.}
\quotateInset{8}{16}{I was just kinda checking on things. Sort of due diligence keeping an eye on the different happenings that the AI was doing at the time.}

\boldify{Last, Paths have an expiration date, meaning that paths not chosen will quickly disappear.
This leads to things being forgotten.
}

Since each event and its corresponding cues were only visible for a limited time, paths not chosen right away by our participants quickly disappeared.
Further, paths are numerous, and frequently updated.
Thus, there is a large risk for paths of inquiry to be forgotten or going unnoticed as the game proceeds, as in these examples:
\quotateInset{7}{14}{Oh my gosh, I didn't even notice he was making an ultralisk den.}
\quotateInset{3}{6}{I didn't notice they canceled the assimilator}

\subsubsection{Many Rapidly Updating Paths: Coping Mechanisms}

\boldify{First (and least interesting) The first was to ``soldier on and ignore it.'' (classic DFS).  This meant the participant paid an INFORMATION cost
}

Our participants responded to this issue in several ways.
First, some participants chose a path and stuck to it, ignoring the others.
Note that this required paying an information cost, because contextual information that may have been very important for future decisions could be discarded in the process.
This strategy was exclusively followed by 3 pairs (2,7,8), who made barely any temporal navigations during the study, as described in Table~\ref{table:taskTime}.
These participants analyzed the replay using not much more time than shoutcasters spend.
However, achieving this speed of analysis required participants to \emph{ignore} many game events.

For example, when asked about desire to click anywhere else, one participant volunteered:
\quotateInset{10}{19}{Mmm, if I had multiple, like, different screens yeah. But no, that seemed to be where the action was gonna be.}
In this fashion, participants chose to triage game events based on some priority order.  
In both of the following examples, the participants navigated away from the conclusion of a fight:
\quotateInset{6}{11}{I wanted to check on his production that one time, because he just lost most of his army, and he still had some [enemies] to deal with.}
\quotateInset{3}{5}{I was trying to see what units they were building, after the fight, see if they were replenishing, or getting ready for another fight.}

\boldify{second coping mechanism we observed is to use the time controls (Rewind/pause/slow down), spending a LONG time on the task as they did a kind of BFS. This meant the participant paid a NAVIGATION cost}

The second method our participants used to manage the complexity of paths was to use the time controls to slow down, stop, or rewind the replay.
Although pausing to assess the state was fairly common in all groups, rewind behavior yielded more information.
Pairs 3 and 10 rewound the most often (Table~\ref{table:taskTime}), and paid higher \emph{navigation} costs to do so, but they viewed these navigations as worthwhile to providing necessary information:
\quotateInset{6}{11}{I looped back to the beginning of the final fight ... to see if there was anything significant that we had missed the first time around.}
However, the cost of doing so was more than just time, because the more paths they monitored, the greater the cognitive load:
\quotateInset{10}{19}{There's just so much happening all at once; I can't keep track of all of it!}

\subsubsection{\ImplicationsText}

\boldify{Implications: we need to improve estimation of value and cost, MANY options are considered, one is chosen.  This contrasts with previous work, where FEW options are considered, but the trouble lies in the DEPTH of the path, as opposed to the NUMBER of paths that must be evaluated. A recommender might be appropriate}

Assessing an agent required considering a great \emph{many} paths, and choosing one (or few), though most paths followed were not particularly \emph{long}.
Note that this contrasts with previous literature in software engineering, which is characterized by ``miles of methods~\cite{piorkowski2016foraging},'' such as a long sequence of methods in the stack trace.
Thus, rapid evaluation and pruning of paths is critical in the RTS domain, but less so in software engineering, where the options to consider are fewer and time pressure is lighter.
One solution could be a recommender system to help the user triage which path to follow next.

\boldify{Implications: Things are being forgotten and paths are not formally concluded often.  Encourage closure of paths, via TODO LISTING}

During assessment, participants' often forgot about or otherwise interrupted their paths of inquiry.
For example, if a new important path appeared, such as a critical battle, either that path or the current path had to be dropped.
In another domain (spreadsheet debugging), participants faced with branching paths with multiple desirable directions became more effective when the environment supported a strategy they call ``to-do listing''~\cite{grigoreanu2010strategy}.  
Because to-do listing was supported on its own or in composition of other problem-solving approaches, it could also act as a strategy enhancer.
Perhaps in the RTS domain, a similar strategy could enable users to carry on with their current path uninterrupted --- but also keep track of the critical battle to come back to later.

%% file: tables/TableWhyCoding.tex
\begin{table}[t]
	\centering
    \small
	\begin{tabular}{@{}p{.89\linewidth}|l@{}}
		\emph{Reasons for participants' path choices code set}&  	
        \textit{Freq}
        \\ \hline
		\ValueWhy{Monitoring State}: Continuous game state monitoring, such as watching a fight.
        \\
        \hspace{.025\linewidth} -\quotate{4}{7}{I wanted to see how the fight was going.}
        & 65
        \\ \hline
		\ValueWhy{New Event}: Attending to a new event for which participant wished to satisfy curiosity about.
        \\
        \hspace{.025\linewidth} -\quotate{2}{4}{I saw there was a new building.}
        & 36
		\\ \hline
		\ValueWhy{Update Game State}: Updating potentially stale game information that the participant explicitly stated prior knowledge about.
        \\
        \hspace{.025\linewidth} -\quotate{1}{2}{I was mainly looking at the army composition, seeing how it had changed from the last fight.}
        & 29
        \\ \hline
        \ValueWhy{Obsolete Domain}: Explicitly using domain info that may not be current, such as game rules (e.g., what buildings can produce).
        \\
        \hspace{.025\linewidth} -\quotate{3}{6}{I mainly clicked on the adept because I'm more familiar with [a previous version of the game].}
        & 11
	\end{tabular}
    \normalsize
	\caption{Reasons for participants' path choices code set, with examples and frequency data, to answer the question ``Why was the participant seeking that information?''
    }
	\label{table:whyCoding}   	
\end{table}

%% file: figures/FigPaths.tex
\begin{figure}
	\centering
    \includegraphics[width=\columnwidth]{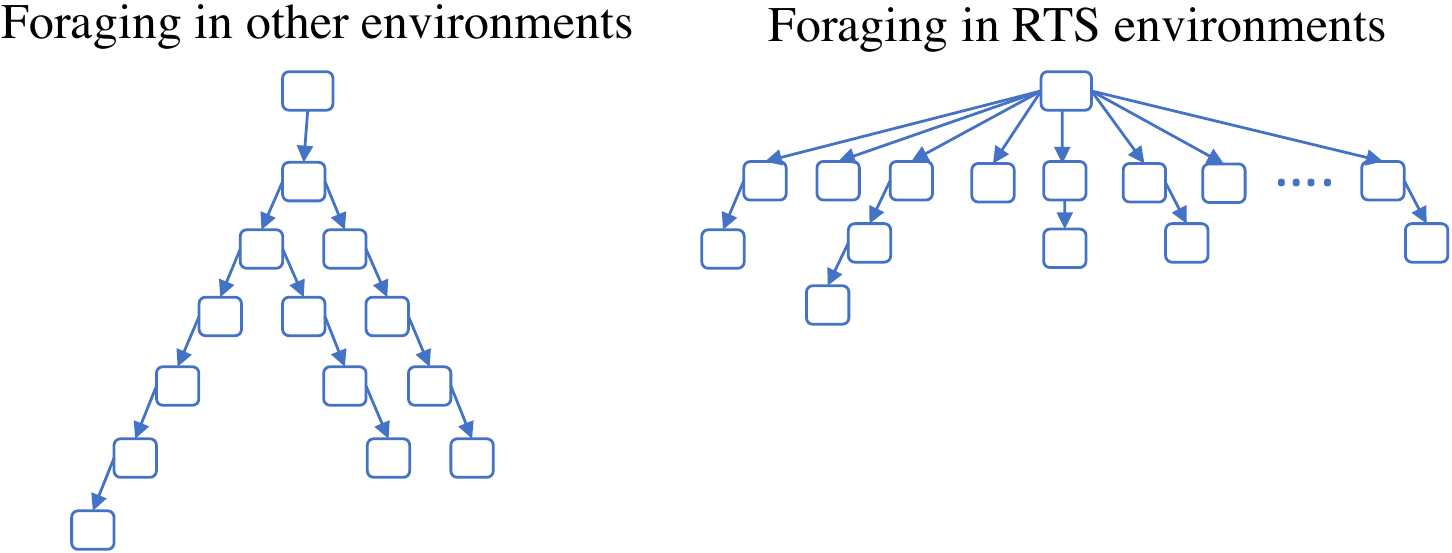}
	\caption{
    Conceptual drawing to contrast foraging in the RTS domain with previously studied foraging.
    (Left): Information environments considered by past IFT literature look like this, where the paths the predator considers are few, but sometimes very deep. 
    This figure is inspired by a programmer's foraging situation in an IDE~[\protect\cite{piorkowski2016foraging} Fig. 5].
    (Right): Foraging in the RTS domain, where most navigation paths are shallow, but with numerous paths to choose from at the top level. 
    }
	\label{fig:paths}
\end{figure}

%% file: tables/TableTaskTime.tex
\begin{table*}[t]
	\centering
    \small
  	\begin{tabular}{@{}l|rr|rr|p{.48\linewidth}@{}}
		
    &	\textit{Task Time}
    & 	\textit{Real-Time Ratio}
    &	\textit{Rewinds}
    &	\textit{Timestamp Rewinds}
    &	\textit{Context Notes}
	\vspace{-12pt}
    \\\hline
    \emph{Pair 1} & 
    20:48 & 1.3
    & 3 & 1 &
    Rewatched 1 fight\\
    \emph{Pair 2} & 20:40 & 1.3
    &   &  &
    Extensive pause around 1:00 to evaluate game state\\
    \emph{Pair 3} & 55:08 & 3.4
    & 12 & 9 &
    Rewatched fights and fight setup. Slowed down replay during 1 combat.\\
    \emph{Pair 4} & 32:16 & 2.0
    & 2 & &
    Rewatched opening build sequence and evaluated information available to the agent at a key moment. Many pauses to explain game state.\\
    \emph{Pair 5} & 24:23 & 1.5
    & 5 & &
    Rewatched unit positioning, AI reaction to events, and scouting effectiveness.\\
    \emph{Pair 6} & 31:56 & 2.0
    & 4 & 2 &
    Rewatched 2 fights.\\
    \emph{Pair 7} & 29:49 & 1.9
    &  & &
    Made no use of time controls other than pausing to write down decision points.\\
    \emph{Pair 8} & 21:27 & 1.3
    & & &
    Made no use of time controls other than pausing to write down decision points.\\
    \emph{Pair 9} & 39:17 & 2.4
    & 2 & 1 &
    Rewatched 1 fight. Slowed down replay for the entire task.\\
    \emph{Pair 10} & 61:14 & 3.8
    & Lots & Some &
    Rewound extensively, in a nested fashion. Changed replay speed many times.\\
\end{tabular}
\normalsize
   \caption{Participant task time (33:42$\pm$14:18 minutes) and time control usage information.
   Note that the replay file was just over 16:04, so dividing each pair's time by 16 yields the third column, ``\emph{Real-Time Ratio}'' (2.1$\pm$0.89).
   Some of times participants rewound the replay were  because we requested timestamps for events, shown in the fourth column, ``\emph{Timestamp Rewinds}.''
   The last column provides any additional context in which replay and pause controls were used. 
   }
	\label{table:taskTime}   	
\end{table*}

%% file: 6-ResultsDecisionPoints.tex
\countedSubsection{RQ3: The Decisions and the Cues}

\boldify{If you're not on the right path, what cue are you following? Let's turn to decision points to figure that out.}

When participants were not heading down the ``right'' path, what cues did they instead follow toward some other path?
Also, what did they consider the ``right'' cues to follow?

\boldify{What do decision points tell us? and what can we do with it}

In the RTS domain, players and intelligent agents make thousands of sequential decisions, and there is a paucity of literature that considers humans trying to understand AI decisions in such a setting.
(A notable exception is McGregor et al.~\cite{mcgregor2015mdp}.)
There is, however, literature that starts with the AI's perspective: instances of its decision-making system components (i.e., neurons) that are interpretable by humans~\cite{Zahavy2016dqn, Zeiler2014visnet}.
In contrast, here we wanted to start with the human's perspective and the foraging paths that result from it: namely, how participants would identify behaviors that were not only potentially human interpretable, but also of interest.

Thus, we asked participants to write down what they thought were the important game events.
We defined the term \emph{key decision points} to our participants as ``an event which is critically important to the outcome of the game,'' to give participants  leeway to apply their own meaning.
Since all participant pairs were examining the same replay file, we were then able to compare the decision points the different participants selected.
That is, the cues in the information environment were the same for all the participants --- whether they noticed them or not.

\boldify{What's the most impt: Expansion!}

Key decision points fell into four main categories: \Ccode{building/producing}, \Ccode{fighting}, \Ccode{moving}, and \Ccode{scouting}.
The participants were in emphatic agreement about the most important types of decision points to pursue.  
Of the 228 total decision points particpants identified, \Ccode{Fighting} and \Ccode{Building} made up 85\% (Table~\ref{table:decisionPointsSummary}).

In fact, participants showed remarkable consistency about the importance of  the \Ccode{Expansion} subcategory of \Ccode{Building}.
Eight of the ten participant pairs identified \Ccode{Expansion} decision points, when a player chooses to build a new resource-producing base (Table~\ref{table:decisionPointsSummary}).
Extra resources from expanding allowed a player to gain an economic advantage over their opponent because they could build more units:
\quotateInset{1}{2}{Of course, if you have a stronger economy you will likely win in the end.}
Moreover, because those that identified any \Ccode{Expansion} found at least three, \Ccode{Expansions} seemed to be considered important throughout the duration of the game.
\quotateInset{6}{11}{... the third base is important for the same reason the first one was, because it was just more production and map presence.}


\input{tables/TableDecisionPointsSummary}

\input{figures/FigDecisionExp}

\boldify{... but some slipped through the cracks.}

Even so, they missed some of the cues pointing out expansion decisions. 
The event logs in the replay file reveal that new bases were constructed at roughly \{1:00, 1:30, 2:00, 5:00, 6:30, 11:20, 12:00, and 13:45\}, each of which is marked with a red line on Figure~\ref{fig:expansionDecisionPoints}.
Only Pair~3 identified decision points for all 8 of these, and 7 pairs omitted at least one\footnote{
Table~\ref{table:decisionPointsSummary} shows Pair 4 also finding eight \Ccode{Expansion} decision points, but one of those is about the \emph{commitment} to expand, based on building other structures to protect the base, rather than the action of building the base itself.}, with one example highlighted with a red box in Figure~\ref{fig:expansionDecisionPoints}.

\boldify{Why did they slip through the cracks? Distractor cues prevented them from noticing at all}

Since \Ccode{Expansion} decisions were so important to our participants, why did they miss some?
``Distractor cues'' in the information environment led participants on other paths\footnote{Reminder: Cues are the signposts in the \emph{environment} that the predator observes, such as rabbit tracks.
Scent, on the other hand, is what the predators make of cues in their \emph{heads}, such as thinking that rabbit tracks will lead to rabbits.}.
Participants were so distracted by cues that provided an alluring scent, albeit to low-value information, they did not notice the other cues that pointed toward the ``Expansion'' decisions. 

Distractor cues led participants astray from \Ccode{Expansion} in nine cases, and eight of them involved units in combat or potentially entering combat.
(The ninth involved being distracted by a scouting unit.)
For example, Pair 7 missed the expansion at the 13:45 minute mark, instead choosing to track various groups of army units, which turned out to be unimportant to them:
\quotateInset{7}{14}{These zerglings are still just chilling.}

\input{figures/FigDecisionFightScout}

\boldify{Distractors were a problem even when the number of interesting events/objects was quite LOW.}

Interestingly, participants had trouble with distractor cues even when the number of events competing for their attention was very low.
For example, in the early stages of the game, players were focused on building economies and scouting.
There was little to no fighting yet, so it was not the source of distracting cues.
We were not surprised that the \Ccode{Expansion} event at 13:45, when the game state had hundreds of objects and events, was the most often missed (5 instances).
However, we were surprised that even when the game state was fairly simple --- such as at 1:30 where the game had only 13 objects --- participants missed the Expansion events. 
The extent of distractibility the partipants showed even when so little was going on was beyond what we expected.

\boldify{So what decision points were noticed after the gamestate gets complex? Fighting!
}

So if decision points went unnoticed in simple game states, what did they notice in complex ones?  
\Ccode{Fighting}.
All participants agreed \Ccode{Fighting} was key, identifying at least one decision point of that type (Table~\ref{table:decisionPointsSummary}).
The ubiquity of \Ccode{Fighting} codes is consistent with Kim et al.~\cite{kim2016evaluation}, who found that combat ratings were the most important to the participant's perception score.
\Ccode{Fighting} provided such a strong scent that it was able to mask most other sources of scent, even those which participants prioritized very highly.

\boldify{Scouting decision points occurred in the first half of the game, but died out once Fighting decision points started occurring in the second half of the game, }

\Ccode{Scouting} offers an example of \Ccode{Fighting} leading participants away from other important patches.
\Ccode{Scouting} decision points occurred in the first half of the game, but died out once \Ccode{Fighting} decision points started to occur in the second half of the game.
As Figure~\ref{fig:decisionFightScout} shows, the start of \Ccode{Fighting} decision points coincides with the time that \Ccode{Scouting} decision points vanish ---
despite the fact that scouting occurred throughout the game, and that participants believed scouting information mattered:
\quotateInset{4}{8}{But it's important just to know what they're up to and good scouting is critical to know who you are going to fight.}

\subsubsection{\ImplicationsText}

\boldify{Fighting is interesting AND important, and is thus very distracting, which can hinder testing the routine, mundane things which continue to be very important.}

Participants had a tendency toward following cues that were  interesting or eye-catching, at the expense of those that were important but more mundane.
In this domain, the ``eye-catching'' cues were combat-oriented, whereas the ``mundane'' cues were scouting oriented.
Other domains may have similar phenomena, wherein certain aspects of the agent's behaviors distract from other important views due to triggering an emotional response in the viewer.
Thus, supporting users' attending to actions that are important but mundane is a design challenge for future interactive explanation systems.

%% file: tables/TableDecisionPointsSummary.tex
\begin{table}[t]
	\setlength{\tabcolsep}{3pt}
	\centering
    \small
	\begin{tabular}{@{}l|l|lllll|lllll@{}}
	     \textit{Code}
         & \emph{Total}
         & \rotatebox{90}{\textit{Pair 1}}
         & \rotatebox{90}{\textit{Pair 2}}
         & \rotatebox{90}{\textit{Pair 3}}
         & \rotatebox{90}{\textit{Pair 4}}
         & \rotatebox{90}{\textit{Pair 5}}
         & \rotatebox{90}{\textit{Pair 6}}
         & \rotatebox{90}{\textit{Pair 7}}
         & \rotatebox{90}{\textit{Pair 8}}
         & \rotatebox{90}{\textit{Pair 9}}
         & \rotatebox{90}{\textit{Pair 10}}
         \\\hline
         
         \Ccode{Expansion} & 52
         & 7
         & 7
         & 8
         & 8
         & -
         & 6
         & 3
         & -
         & 7
         & 6 
         \\
         
         \Ccode{Building} - Rest & 69
         & 6
         & 4
         & 7
         & 15
         & 1
         & 7
         & 11
         & 2
         & 12
         & 4 
         \\\hline
         
         \Ccode{Building} - All & 114
         & 13
         & 11
         & 15
         & 21
         & 1
         & 12
         & 12
         & 2
         & 18
         & 9
         \\ 
   
         \Ccode{Fighting} - All & 98
         & 8
         & 4
         & 11
         & 8
         & 4
         & 10
         & 6
         & 8
         & 11
         & 28
		 \\
         
         \Ccode{Moving} - All & 26
         & 3
         & 1
         & 5
         & 1
         & 2
         & 2
         & 2
         & 2
         & 3
         & 5 
         \\
         
         \Ccode{Scouting} - All & 23
         & 1
         & 2
         & 1
         & 5
         & 1
         & 1
         & 3
         & -
         & 3
         & 6 
         \\\hline  \hline

         {\hfill\textit{Total}} & 228
         & 34
         & 20
         & 40
         & 45
         & 11
         & 31
         & 39
         & 16
         & 42
         & 65 \\

        \end{tabular}
        \normalsize
	\caption{Summary of decision points identified by our participants. 
    Sums may exceed totals, since each decision point could have multiple labels.
    Note how prevalent \Ccode{Expansion} was within the \Ccode{Building} category.
   }
    
	\label{table:decisionPointsSummary}   	
\end{table}

%% file: figures/FigDecisionExp.tex
\begin{figure}
	\centering
	\includegraphics[width=\linewidth]{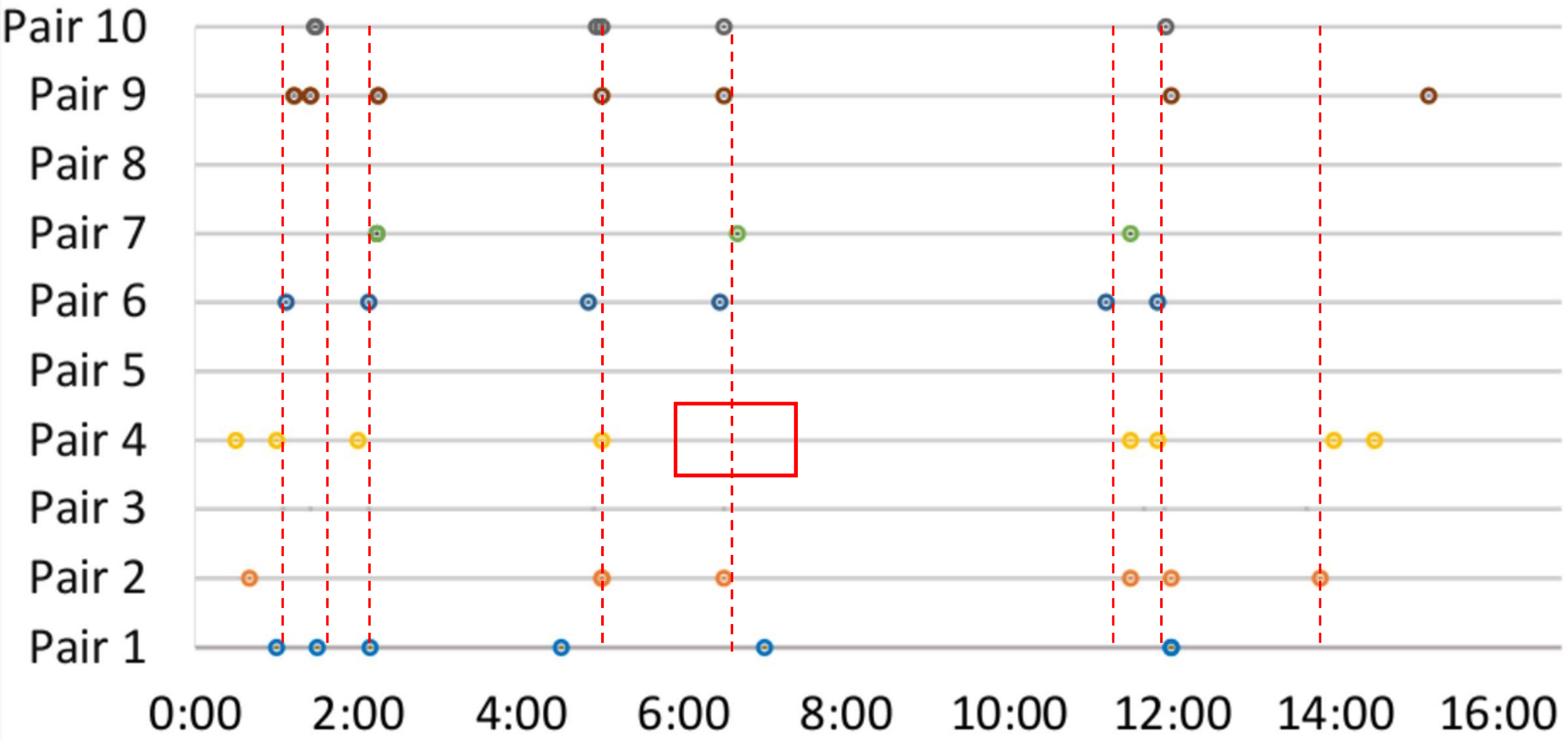}
	\caption{
    All \Ccode{Building-Expansion} decision points identified by our participants (y-axis), with game time on the x-axis.
    \Ccode{Expansion} events are known to have occurred in the replay file at roughly: \{1:00, 1:30, 2:00, 5:00, 6:30, 11:20, 12:00, and 13:45\}.
    Each of these times is demarcated on the figure with a red vertical line, often coinciding with decision points.
    Consider the red box, where Pair 4 failed to notice an event they likely wanted to note, based on their previous and subsequent behavior.
}
	\label{fig:expansionDecisionPoints}
\end{figure}

%% file: figures/FigDecisionFightScout.tex
\begin{figure}
	\centering
    \includegraphics[width=\columnwidth]
{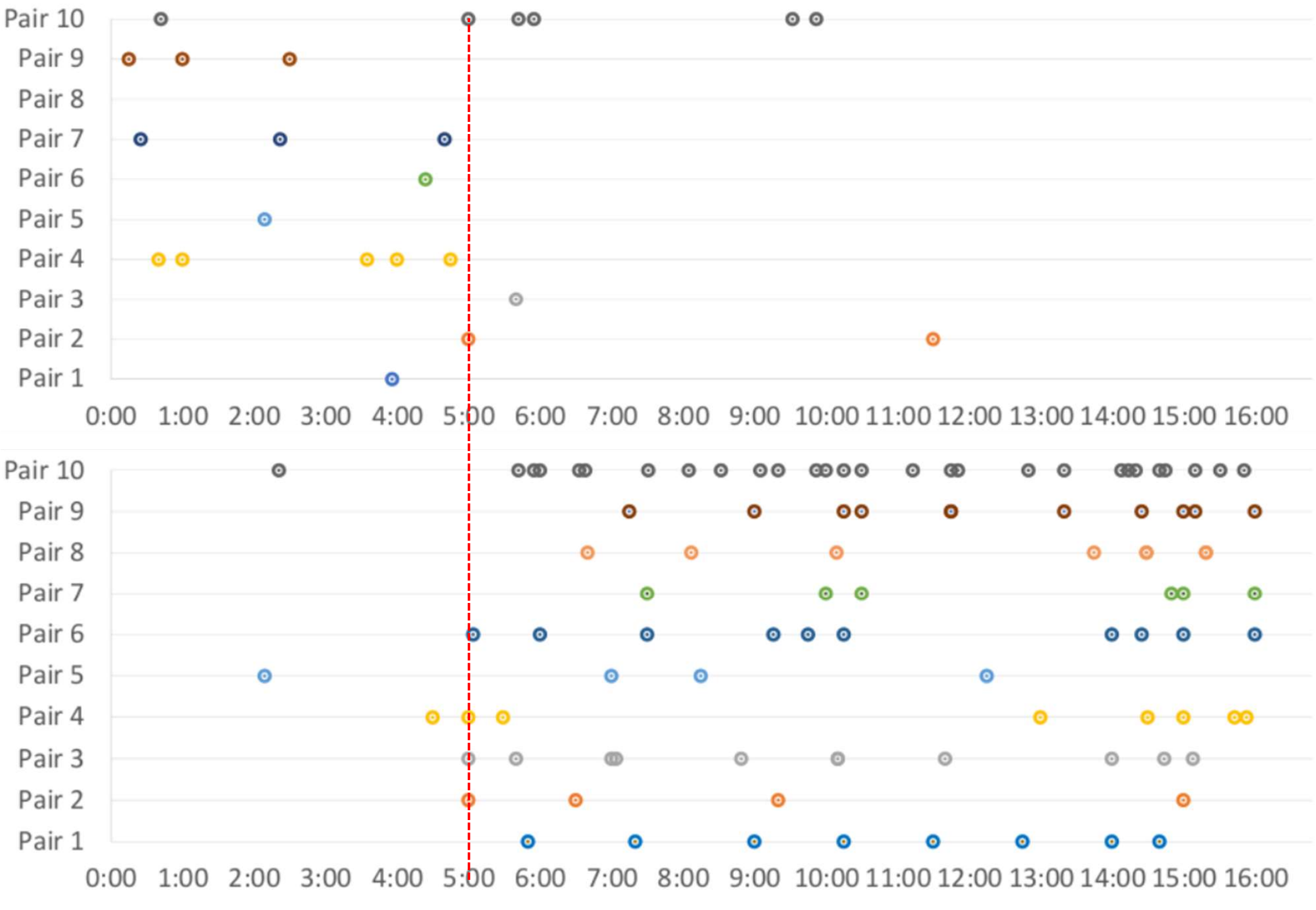}
	\caption{
    (Top:) All \Ccode{Scouting} decision points identified by our participant pairs (y-axis), with game time on the x-axis.
    (Bottom:) All \Ccode{Fighting} decision points identified, plotted on the same axes. 
    The red line that passes through both images denotes roughly the time at which \Ccode{Fighting} events begin.
    Notice that after this time, many \Ccode{Fighting} decision points are identified, but \Ccode{Scouting} decision points are no longer noticed often -- despite important \Ccode{Scouting} actions continuing to occur.
}

	\label{fig:decisionFightScout}
\end{figure}

%% file: 8-Discussion.tex
\countedSection{Threats to Validity}
\boldify{Every study has threats to validity.  Among ours were...}

Every study has threats to validity~\cite{Wohlin-2012}. 
This paper presents the first study of information foraging either in the area of explainable AI or in the domain of RTS games, so its results cannot yet be compared or validated by other studies by other researchers.
Thus, we must be conscious of its limitations.

Aspects of our study may have influenced our participants to ask less questions in general, such as not asking a question of their partner if they did not expect their partner to be able to answer it.
Also, participants took different amounts of time to do the task, ranging from 20 minutes to an hour.
Thus, certain participant pairs talked more than others in the main task, creating a form of sampling bias.
Threats like these can be addressed only by additional
empirical studies across a spectrum of study designs, types of intelligent interfaces, and intelligent agents.

\countedSection{Discussion: What IFT Can Offer Explainable AI}

\boldify{In this paper, we've used an IFT perspective, which enabled us to abstract above game-specific puzzlements to constructs grounded in a well-established theory of humans doing information seeking: namely what prey users sought (RQ1), the paths they took to get there (RQ2), and the cues that they used to select navigations (RQ3). }

At this point, we step back to consider insights an Information Foraging Theory perspective can bring to Explainable AI.

Perhaps most important, the theory allows us to ``connect the dots'' between our work and other work done from an IFT perspective.
It does so by enabling us to abstract beyond game-specific puzzlements to constructs grounded in a well-established theory for humans' information seeking behaviors.

Thus, we used IFT to abstract above game objects like ``assimilators'' to the IFT constructs of \emph{prey} (RQ1), foraging along \emph{paths} (RQ2), and why they followed the \emph{cues} they followed (RQ3).
The IFT lens revealed that participants faced difficult foraging problems -- some of which are new to IFT research -- 
and faced high foraging costs.
For example, failure to follow the ``right'' paths resulted in a high \emph{information cost} being paid, but finding a reasonable path needed to be done quickly due to the ever-changing game environment (at a high \emph{cognitive cost}).
Although the user could relax the real-time pressure by pausing the replay, excessive rewinding incurred not only a high \emph{navigation cost} for rewind-positioning and pausing, but also an additional \emph{cognitive cost} of remembering more context.

\boldify{Navigation-information Cost tradeoff: This leads to a type of cognitive cost stemming from evaluating and tracking many paths if the participant chose to take multiple paths OR a type of cognitive cost stemming from loss of contextual information if the participant chose to triage}

Participants had to make trade offs between two types of these costs, navigational and informational, and their triaging to manage such trade-offs led to even more cognitive cost.
Each path participants followed incurred a navigational cost, so following more paths led to higher cumulative costs.
However, reducing the number of paths they followed incurred the information cost of missing out on potentially important information.
Worse, the information cost paid by adhering to a single path compounded over time.
For example, if participants made a bad path choice early in the game, and repeated that mistake throughout the game, then later in the game they may be confused by an event that they did not expect --- due to lacking appropriate context.
One cause for making a bad choice was ``distractor cues,'' where, in order to curtail their current navigation direction and move to another, participants paid the high information cost of missing information important to them, often unwittingly.

\boldify{The prey is necessarily in pieces, both in how we framed the problem and how the participants solved it.}

The IFT perspective also connects some of the problems our participants faced to known problems of foraging in other domains.
One open problem in IFT is the Prey in Pieces problem\footnote{Piorkowski et al. described ``Prey in Pieces'' as if getting a coffeemaker meant a shopper had to buy individual parts at different stores, then finally piece them together.  
The cost of going to every store must be paid plus the cost of piecing things together at the end, rather than the cost of going to one store that has a preassembled coffeemaker.}~\cite{piorkowski2016foraging}.
Our participants encountered this problem because they had to piece together bits of evidence of the agent's decisions in order to assess the agent.
In doing so, participants were sometimes uncertain about what each of these decisions meant about the competencies and strategies of the agent. 
When aggregating multiple sources of uncertain data like this, prior research has shown that computational assistance can increase user confidence, although manual comparison is still preferred in high-stakes situations~\cite{Greis2017uncertain}.
This seems to suggest that a recommender may be helpful to help users  
select and aggregate agent actions for explanation, though manual comparison may still be necessary at times.

\FIXME{JED: FIXED? edited last sentence in response to Sruti's feedback on whether stakes were low or high, and what the citation means}

\boldify{The Scaling Up problem in IFT is different here, compared to other domains. In the value and cost (2016 fse) paper, the scaling up problem referred to the difficulties programmers faced with estimating the cost and value of patches more than one click away.  This is a depth problem.  What we generally saw was a breadth problem, and our users need to know which of these many paths to choose from to take in order to maximize V/C}

Another open problem in IFT is the Scaling Up problem~\cite{piorkowski2016foraging}.
This problem was revealed in the domain of IDEs, in which foragers (developers) had great difficulty accurately predicting the cost and value of going to patches more than one link away.
The problem that the developers faced was a \emph{depth} problem (recall Figure~\ref{fig:paths}).
In contrast, in our domain, participants faced a \emph{breadth} Scaling Up foraging problem: constantly having to choose which of many paths to follow.
The Scaling Up problem as a depth problem is still open; so too is the breadth version of it identified here.

In both cases, users foraging for information want to maximize value per cost. 
In the IDE case, this is accomplished by pruning low value paths unrelated to the bug. 
For example, if a developer is fixing a UI bug, they can potentially ignore database code. 
However, in RTS, any action could be important, so many paths need to stay on the table. 
Further, the rapid rate of change in the environment limits the user's planning depth, which decreases accuracy of predictions of cost/value.
Thus, the Scaling Up problem is different in the RTS domain.
In depth domains like IDEs, the problem is predicting cost/value in far-distant patches,
whereas in the RTS domain, the difficulty is rapidly choosing at the top level among the many, many available paths.

%% file: 9-Conclusion.tex
\countedSection{Conclusion}

In this paper, we presented the first theory-based investigation into how people forage for information about an intelligent agent in an RTS environment and the  implications for Explainable AI. 
Our results suggest that people's information seeking in this domain is far from straightforward. 
We saw evidence of this from multiple perspectives:

\begin{enumerate}[labelindent=20pt,labelwidth=\widthof{\ref{last-item-conclusion}},label=\arabic*.,itemindent=1em,leftmargin=!]
\item[\textbf{RQ1}] \emph{The Prey}: Participants favored \LimDey{What} information over the \LimDey{Why}s reported by most previous research, and their \LimDey{What}s were nuanced, complex, and sometimes expensive. 
\vspace{-5pt}
\item[\textbf{RQ2}] \emph{The Paths}: The dynamically changing RTS environment and the breadth-oriented structure of its information paths caused unique information foraging problems in deciding which paths to traverse. These problems led not only to navigation costs, but also to information and  cognitive costs.
\vspace{-5pt}
\item[\textbf{RQ3}] \emph{The Decisions and the Cues}: These costs rendered it infeasible for participants to investigate all of the decision points they wanted. This problem was exacerbated by ``distractor cues,'' which drew participants' attention elsewhere with interesting-looking cues (like signs of fighting), at the expense of information that was often important to participants (like scouting or expansion).\label{last-item-conclusion} 

\end{enumerate}

Perhaps most importantly, our results point to the benefits of investigating humans' understanding of intelligent agents through the lens of Information Foraging Theory. 
For example, the IFT lens enabled us to abstract beyond StarCraft, to reveal phenomena -- such as the frequent need to trade off cognitive foraging costs against navigation foraging costs against information costs -- that are widely relevant to the RTS domain.
As we have noted in the ``Implications'' sections along the way, 
these theory-based results reveal opportunities for future Explainable AI systems to enable domain experts to find the information they need to understand, assess, and ultimately decide how much to trust their intelligent agents.

%% file: 0-paper.bbl

\begin{thebibliography}{00}


\ifx \showCODEN    \undefined \def \showCODEN     #1{\unskip}     \fi
\ifx \showDOI      \undefined \def \showDOI       #1{{\tt DOI:}\penalty0{#1}\ }
  \fi
\ifx \showISBNx    \undefined \def \showISBNx     #1{\unskip}     \fi
\ifx \showISBNxiii \undefined \def \showISBNxiii  #1{\unskip}     \fi
\ifx \showISSN     \undefined \def \showISSN      #1{\unskip}     \fi
\ifx \showLCCN     \undefined \def \showLCCN      #1{\unskip}     \fi
\ifx \shownote     \undefined \def \shownote      #1{#1}          \fi
\ifx \showarticletitle \undefined \def \showarticletitle #1{#1}   \fi
\ifx \showURL      \undefined \def \showURL       #1{#1}          \fi

\bibitem{beltran2017don}
{Juan~Felipe Beltran}, {Ziqi Huang}, {Azza Abouzied}, {and} {Arnab Nandi}.
  2017.
\newblock \showarticletitle{Don't just swipe left, tell me why: Enhancing
  gesture-based feedback with reason bins}. In {\em Proceedings of the 22nd
  International Conference on Intelligent User Interfaces}. ACM, 469--480.
\newblock


\bibitem{bhowmick2013not}
{Sourav~S Bhowmick}, {Aixin Sun}, {and} {Ba~Quan Truong}. 2013.
\newblock \showarticletitle{Why Not, WINE?: Towards answering why-not questions
  in social image search}. In {\em Proceedings of the 21st ACM International
  Conference on Multimedia}. ACM, 917--926.
\newblock


\bibitem{bostandjiev2012tasteweights}
{Svetlin Bostandjiev}, {John O'Donovan}, {and} {Tobias H{\"o}llerer}. 2012.
\newblock \showarticletitle{TasteWeights: a visual interactive hybrid
  recommender system}. In {\em Proceedings of the Sixth ACM Conference on
  Recommender Systems}. ACM, 35--42.
\newblock


\bibitem{castelli2017happened}
{Nico Castelli}, {Corinna Ogonowski}, {Timo Jakobi}, {Martin Stein}, {Gunnar
  Stevens}, {and} {Volker Wulf}. 2017.
\newblock \showarticletitle{What Happened in my home? An end-user development
  approach for smart home data visualization}. In {\em Proceedings of the 2017
  CHI Conference on Human Factors in Computing Systems}. ACM, 853--866.
\newblock


\bibitem{Cheung:2011:SSU:1978942.1979053}
{Gifford Cheung} {and} {Jeff Huang}. 2011.
\newblock \showarticletitle{Starcraft from the Stands: Understanding the Game
  Spectator}. In {\em Proceedings of the SIGCHI Conference on Human Factors in
  Computing Systems} {\em (CHI '11)}. ACM, New York, NY, USA, 763--772.
\newblock
\showISBNx{978-1-4503-0228-9}
\showDOI{%
\url{http://dx.doi.org/10.1145/1978942.1979053}}


\bibitem{chi2001using}
{Ed~H Chi}, {Peter Pirolli}, {Kim Chen}, {and} {James Pitkow}. 2001.
\newblock \showarticletitle{Using information scent to model user information
  needs and actions and the web}. In {\em Proceedings of the SIGCHI Conference
  on Human Factors in Computing Systems}. ACM, 490--497.
\newblock


\bibitem{cotter2017explaining}
{Kelley Cotter}, {Janghee Cho}, {and} {Emilee Rader}. 2017.
\newblock \showarticletitle{Explaining the news feed algorithm: An analysis of
  the}. In {\em Proceedings of the 2017 CHI Conference Extended Abstracts on
  Human Factors in Computing Systems}. ACM, 1553--1560.
\newblock


\bibitem{fleming2013information}
{Scott~D. Fleming}, {Chris Scaffidi}, {David Piorkowski}, {Margaret Burnett},
  {Rachel Bellamy}, {Joseph Lawrance}, {and} {Irwin Kwan}. 2013.
\newblock \showarticletitle{An information foraging theory perspective on tools
  for debugging, refactoring, and reuse tasks}.
\newblock {\em ACM Transactions on Software Engineering and Methodology
  (TOSEM)\/} {22}, 2 (2013), 14.
\newblock


\bibitem{fu2007snif}
{Wai-Tat Fu} {and} {Peter Pirolli}. 2007.
\newblock \showarticletitle{SNIF-ACT: A cognitive model of user navigation on
  the world wide web}.
\newblock {\em Human-Computer Interaction\/} {22}, 4 (2007), 355--412.
\newblock


\bibitem{Greis2017uncertain}
{Miriam Greis}, {Emre Avci}, {Albrecht Schmidt}, {and} {Tonja Machulla}. 2017.
\newblock \showarticletitle{Increasing users' confidence in uncertain data by
  aggregating data from multiple sources}. In {\em Proceedings of the 2017 CHI
  Conference on Human Factors in Computing Systems} {\em (CHI '17)}. ACM, New
  York, NY, USA, 828--840.
\newblock
\showISBNx{978-1-4503-4655-9}
\showDOI{%
\url{http://dx.doi.org/10.1145/3025453.3025998}}


\bibitem{grigoreanu2010strategy}
{Valentina~I Grigoreanu}, {Margaret~M Burnett}, {and} {George~G Robertson}.
  2010.
\newblock \showarticletitle{A strategy-centric approach to the design of
  end-user debugging tools}. In {\em Proceedings of the SIGCHI Conference on
  Human Factors in Computing Systems}. ACM, 713--722.
\newblock


\bibitem{hayes2017improving}
{Bradley Hayes} {and} {Julie~A Shah}. 2017.
\newblock \showarticletitle{Improving robot controller transparency through
  autonomous policy explanation}. In {\em Proceedings of the 2017 ACM/IEEE
  International Conference on Human-Robot Interaction}. ACM, 303--312.
\newblock


\bibitem{he2014answering}
{Zhian He} {and} {Eric Lo}. 2014.
\newblock \showarticletitle{Answering why-not questions on top-k queries}.
\newblock {\em IEEE Transactions on Knowledge and Data Engineering\/} {26}, 6
  (2014), 1300--1315.
\newblock


\bibitem{kapoor2010interactive}
{Ashish Kapoor}, {Bongshin Lee}, {Desney Tan}, {and} {Eric Horvitz}. 2010.
\newblock \showarticletitle{Interactive optimization for steering machine
  classification}. In {\em Proceedings of the SIGCHI Conference on Human
  Factors in Computing Systems}. ACM, 1343--1352.
\newblock


\bibitem{kim2016evaluation}
{Man-Je Kim}, {Kyung-Joong Kim}, {SeungJun Kim}, {and} {Anind~K Dey}. 2016.
\newblock \showarticletitle{Evaluation of StarCraft Artificial Intelligence
  Competition Bots by Experienced Human Players}. In {\em Proceedings of the
  2016 CHI Conference Extended Abstracts on Human Factors in Computing
  Systems}. ACM, 1915--1921.
\newblock


\bibitem{kulesza2015principles}
{Todd Kulesza}, {Margaret Burnett}, {Weng-Keen Wong}, {and} {Simone Stumpf}.
  2015.
\newblock \showarticletitle{Principles of explanatory debugging to personalize
  interactive machine learning}. In {\em Proceedings of the 20th International
  Conference on Intelligent User Interfaces}. ACM, 126--137.
\newblock


\bibitem{kulesza2012tell}
{Todd Kulesza}, {Simone Stumpf}, {Margaret Burnett}, {and} {Irwin Kwan}. 2012.
\newblock \showarticletitle{Tell me more? The effects of mental model soundness
  on personalizing an intelligent agent}. In {\em Proceedings of the SIGCHI
  Conference on Human Factors in Computing Systems}. ACM, 1--10.
\newblock


\bibitem{kulesza2010explanatory}
{Todd Kulesza}, {Simone Stumpf}, {Margaret Burnett}, {Weng-Keen Wong}, {Yann
  Riche}, {Travis Moore}, {Ian Oberst}, {Amber Shinsel}, {and} {Kevin
  McIntosh}. 2010.
\newblock \showarticletitle{Explanatory debugging: Supporting end-user
  debugging of machine-learned programs}. In {\em Visual Languages and
  Human-Centric Computing (VL/HCC), 2010 IEEE Symposium on}. IEEE, 41--48.
\newblock


\bibitem{kulesza2011oriented}
{Todd Kulesza}, {Simone Stumpf}, {Weng-Keen Wong}, {Margaret~M Burnett},
  {Stephen Perona}, {Andrew Ko}, {and} {Ian Oberst}. 2011.
\newblock \showarticletitle{Why-oriented end-user debugging of naive Bayes text
  classification}.
\newblock {\em ACM Transactions on Interactive Intelligent Systems (TiiS)\/}
  {1}, 1 (2011), 2.
\newblock


\bibitem{kuttal2013predator}
{Sandeep~Kaur Kuttal}, {Anita Sarma}, {and} {Gregg Rothermel}. 2013.
\newblock \showarticletitle{Predator behavior in the wild web world of bugs: An
  information foraging theory perspective}. In {\em Visual Languages and
  Human-Centric Computing (VL/HCC), 2013 IEEE Symposium on}. IEEE, 59--66.
\newblock


\bibitem{lim2009assessing}
{Brian~Y Lim} {and} {Anind~K Dey}. 2009.
\newblock \showarticletitle{Assessing demand for intelligibility in
  context-aware applications}. In {\em Proceedings of the 11th International
  Conference on Ubiquitous Computing}. ACM, 195--204.
\newblock


\bibitem{lim2009}
{Brian~Y. Lim}, {Anind~K. Dey}, {and} {Daniel Avrahami}. 2009.
\newblock \showarticletitle{Why and why not explanations improve the
  intelligibility of context-aware intelligent systems}. In {\em Proceedings of
  the SIGCHI Conference on Human Factors in Computing Systems}. ACM,
  2119--2128.
\newblock


\bibitem{lomas2012robots}
{M. Lomas}, {R. Chevalier}, {E.~V. Cross}, {R.~C. Garrett}, {J. Hoare}, {and}
  {M. Kopack}. 2012.
\newblock \showarticletitle{Explaining robot actions}. In {\em 2012 7th
  ACM/IEEE International Conference on Human-Robot Interaction (HRI)}.
  187--188.
\newblock
\showISSN{2167-2121}
\showDOI{%
\url{http://dx.doi.org/10.1145/2157689.2157748}}


\bibitem{mcgregor2015mdp}
{S. McGregor}, {H. Buckingham}, {T.~G. Dietterich}, {R. Houtman}, {C.
  Montgomery}, {and} {R. Metoyer}. 2015.
\newblock \showarticletitle{Facilitating testing and debugging of Markov
  Decision Processes with interactive visualization}. In {\em 2015 IEEE
  Symposium on Visual Languages and Human-Centric Computing (VL/HCC)}. 53--61.
\newblock
\showDOI{%
\url{http://dx.doi.org/10.1109/VLHCC.2015.7357198}}


\bibitem{metoyer2010explaining}
{Ronald Metoyer}, {Simone Stumpf}, {Christoph Neumann}, {Jonathan Dodge}, {Jill
  Cao}, {and} {Aaron Schnabel}. 2010.
\newblock \showarticletitle{Explaining how to play real-time strategy games}.
\newblock {\em Knowledge-Based Systems\/} {23}, 4 (2010), 295--301.
\newblock


\bibitem{miller2017social}
{Tim Miller}. 2017.
\newblock \showarticletitle{Explanation in Artificial Intelligence: Insights
  from the Social Sciences}.
\newblock {\em CoRR\/}  {abs/1706.07269} (2017).
\newblock
\showURL{%
\url{http://arxiv.org/abs/1706.07269}}


\bibitem{niu2013departures}
{Nan Niu}, {Anas Mahmoud}, {Zhangji Chen}, {and} {Gary Bradshaw}. 2013.
\newblock \showarticletitle{Departures from optimality: Understanding human
  analyst's information foraging in assisted requirements tracing}. In {\em
  Proceedings of the 2013 International Conference on Software Engineering}.
  IEEE Press, 572--581.
\newblock


\bibitem{norman1983some}
{Donald~A Norman}. 1983.
\newblock \showarticletitle{Some observations on mental models}.
\newblock {\em Mental Models\/} {7}, 112 (1983), 7--14.
\newblock


\bibitem{ontanon}
{S. Onta{\~n}{\'o}n}, {G. Synnaeve}, {A. Uriarte}, {F. Richoux}, {D.
  Churchill}, {and} {M. Preuss}. 2013.
\newblock \showarticletitle{A Survey of Real-Time Strategy Game AI Research and
  Competition in StarCraft}.
\newblock {\em IEEE Transactions on Computational Intelligence and AI in
  Games\/} {5}, 4 (Dec 2013), 293--311.
\newblock
\showISSN{1943-068X}
\showDOI{%
\url{http://dx.doi.org/10.1109/TCIAIG.2013.2286295}}


\bibitem{perez2014diagnosis}
{Alexandre Perez} {and} {Rui Abreu}. 2014.
\newblock \showarticletitle{A diagnosis-based approach to software
  comprehension}. In {\em Proceedings of the 22nd International Conference on
  Program Comprehension}. ACM, 37--47.
\newblock


\bibitem{piorkowski2015fix}
{David Piorkowski}, {Scott~D. Fleming}, {Christopher Scaffidi}, {Margaret
  Burnett}, {Irwin Kwan}, {Austin~Z Henley}, {Jamie Macbeth}, {Charles Hill},
  {and} {Amber Horvath}. 2015.
\newblock \showarticletitle{To fix or to learn? How production bias affects
  developers' information foraging during debugging}. In {\em Software
  Maintenance and Evolution (ICSME), 2015 IEEE International Conference on}.
  IEEE, 11--20.
\newblock


\bibitem{piorkowski2016foraging}
{David Piorkowski}, {Austin~Z Henley}, {Tahmid Nabi}, {Scott~D Fleming},
  {Christopher Scaffidi}, {and} {Margaret Burnett}. 2016.
\newblock \showarticletitle{Foraging and navigations, fundamentally:
  developers' predictions of value and cost}. In {\em Proceedings of the 2016
  24th ACM SIGSOFT International Symposium on Foundations of Software
  Engineering}. ACM, 97--108.
\newblock


\bibitem{pirolli2007information}
{Peter Pirolli}. 2007.
\newblock {\em Information Foraging Theory: Adaptive Interaction with
  Information}.
\newblock Oxford University Press.
\newblock


\bibitem{Rosenthal2016robots}
{Stephanie Rosenthal}, {Sai~P. Selvaraj}, {and} {Manuela Veloso}. 2016.
\newblock \showarticletitle{Verbalization: Narration of autonomous robot
  experience}. In {\em Proceedings of the Twenty-Fifth International Joint
  Conference on Artificial Intelligence} {\em (IJCAI'16)}. AAAI Press,
  862--868.
\newblock
\showISBNx{978-1-57735-770-4}
\showURL{%
\url{http://dl.acm.org/citation.cfm?id=3060621.3060741}}


\bibitem{srinivasa2016foraging}
{Sruti Srinivasa~Ragavan}, {Sandeep~Kaur Kuttal}, {Charles Hill}, {Anita
  Sarma}, {David Piorkowski}, {and} {Margaret Burnett}. 2016.
\newblock \showarticletitle{Foraging among an overabundance of similar
  variants}. In {\em Proceedings of the 2016 CHI Conference on Human Factors in
  Computing Systems}. ACM, 3509--3521.
\newblock


\bibitem{stumpf2007}
{Simone Stumpf}, {Vidya Rajaram}, {Lida Li}, {Margaret Burnett}, {Thomas
  Dietterich}, {Erin Sullivan}, {Russell Drummond}, {and} {Jonathan Herlocker}.
  2007.
\newblock \showarticletitle{Toward harnessing user feedback for machine
  learning}. In {\em Proceedings of the 12th International Conference on
  Intelligent user interfaces}. ACM, 82--91.
\newblock


\bibitem{sycara2015abstraction}
{Katia Sycara}, {Christian Lebiere}, {Yulong Pei}, {Donald Morrison}, {and}
  {Michael Lewis}. 2015.
\newblock \showarticletitle{Abstraction of analytical models from cognitive
  models of human control of robotic swarms}. In {\em International Conference
  on Cognitive Modeling}. University of Pittsburgh.
\newblock


\bibitem{tullio2007}
{Joe Tullio}, {Anind~K Dey}, {Jason Chalecki}, {and} {James Fogarty}. 2007.
\newblock \showarticletitle{How it works: A field study of non-technical users
  interacting with an intelligent system}. In {\em Proceedings of the SIGCHI
  Conference on Human Factors in Computing Systems}. ACM, 31--40.
\newblock


\bibitem{vermeulen2010pervasivecrystal}
{Jo Vermeulen}, {Geert Vanderhulst}, {Kris Luyten}, {and} {Karin Coninx}. 2010.
\newblock \showarticletitle{PervasiveCrystal: Asking and answering why and why
  not questions about pervasive computing applications}. In {\em Intelligent
  Environments (IE), 2010 Sixth International Conference on}. IEEE, 271--276.
\newblock


\bibitem{vinyals}
{Oriol Vinyals}. 2017.
\newblock DeepMind and Blizzard open StarCraft II as an AI research
  environment.
\newblock   (2017).
\newblock
\showURL{%
\url{https://deepmind.com/blog/deepmind-and-blizzard-open-starcraft-ii-ai-research-environment/}}


\bibitem{Wohlin-2012}
{Claes Wohlin}, {Per Runeson}, {Martin H\"{o}st}, {Magnus~C. Ohlsson},
  {Bj\"{o}orn Regnell}, {and} {Anders Wessl{\'e}n}. 2000.
\newblock {\em Experimentation in Software Engineering: An Introduction}.
\newblock Kluwer Academic Publishers, Norwell, MA, USA.
\newblock
\showISBNx{0-7923-8682-5}


\bibitem{wong_2016}
{Kevin Wong}. 2016.
\newblock StarCraft 2 and the quest for the highest APM.
\newblock   (Jul 2016).
\newblock
\showURL{%
\url{https://www.engadget.com/2014/10/24/starcraft-2-and-the-quest-for-the-highest-apm/}}


\bibitem{Zahavy2016dqn}
{Tom Zahavy}, {Nir~Ben Zrihem}, {and} {Shie Mannor}. 2016.
\newblock \showarticletitle{Graying the black box: Understanding DQNs}. In {\em
  Proceedings of the 33rd International Conference on International Conference
  on Machine Learning - Volume 48} {\em (ICML'16)}. JMLR.org, 1899--1908.
\newblock
\showURL{%
\url{http://dl.acm.org/citation.cfm?id=3045390.3045591}}


\bibitem{Zeiler2014visnet}
{Matthew~D. Zeiler} {and} {Rob Fergus}. 2014.
\newblock {\em Visualizing and understanding convolutional networks}.
\newblock Springer International Publishing, Cham, 818--833.
\newblock
\showISBNx{978-3-319-10590-1}
\showDOI{%
\url{http://dx.doi.org/10.1007/978-3-319-10590-1_53}}


\end{thebibliography}
